%% file: main.tex
\documentclass[acmlarge,screen]{acmart}

\usepackage{tikz}
\usepackage[english]{babel}
\usepackage{moresize}
\usepackage{amsmath}

\usepackage{balance}
\usepackage{wrapfig}
\usepackage{comment}
\usepackage{paralist}
\usepackage{multirow}
\usepackage{pgfplots}

\usepackage{filecontents}

\usepackage[english]{babel}
\usepackage[latin1]{inputenc}
\usepackage{mathrsfs}
\usepackage{graphicx}
\usepackage{amssymb}
\usepackage{url}
\usepackage{subfigure}
\usepackage{amsmath}
\usepackage{enumitem}
\usepackage[linesnumbered,algoruled,boxed,lined]{algorithm2e}
\usepackage{adjustbox}
\usepackage{amssymb}
\usepackage{times}
\usepackage{hyperref}
\usepackage{filecontents}
\definecolor{tblue}{RGB}{31,119,180}
\definecolor{torange}{RGB}{255,127,14}
\definecolor{tgreen}{RGB}{44,160,44}
\definecolor{tred}{RGB}{214,39,40}
\definecolor{tpurple}{RGB}{148,103,189}

\usepackage{filecontents}

\newcommand{\hide}[1]{} 

\newcommand{\etal}{\textit{et al}.}

\newcommand{\ie}{\textit{i}.\textit{e}.}
\newcommand{\eg}{\textit{e}.\textit{g}.} 
\newcommand{\wrt}{\textit{w}.\textit{r}.\textit{t}} 
\newtheorem{Dfn}{Definition}

\newcommand{\model}{\textit{HeartSpace}} 

\AtBeginDocument{%
  \providecommand\BibTeX{{%
    \normalfont B\kern-0.5em{\scshape i\kern-0.25em b}\kern-0.8em\TeX}}}

\setcopyright{none}
\renewcommand\footnotetextcopyrightpermission[1]{}

\begin{document}

\title{Representation Learning on Variable Length and Incomplete Wearable-Sensory Time Series}

\author{Xian Wu}
\email{xwu9@nd.edu}
\author{Chao Huang}
\author{Pablo Roblesgranda}
\author{Nitesh Chawla}
\email{nchawla@nd.edu}
\affiliation{%
  \institution{Department of Computer Science and Engineering, University of Notre Dame}
  \city{Notre Dame}
  \state{IN}
  \postcode{46556}
}

\renewcommand{\shortauthors}{Xian Wu, et al.}

\begin{abstract}
The prevalence of wearable sensors (\eg, smart wristband) is creating unprecedented opportunities to not only inform health and wellness states of individuals, but also assess and infer personal attributes, including demographic and personality attributes. However, the data captured from wearables, such as heart rate or number of steps, present two key challenges: 1) the time series is often of variable-length and incomplete due to different data collection periods (\eg, wearing behavior varies by person); and 2) inter-individual variability to external factors like stress and environment. \hide{3) inter-individual variability to external factors like stress and environment; and 4) lack of substantial ground truth.} This paper addresses these challenges and brings us closer to the potential of personalized insights about an individual, taking the leap from quantified self to qualified self. 
Specifically, \emph{\model} proposed in this paper encodes time series data with variable-length and missing values via the integration of a time series encoding module and a pattern aggregation network. Additionally, \emph{\model} implements a Siamese-triplet network to optimize representations by jointly capturing intra- and inter-series correlations during the embedding learning process. The empirical evaluation over two different real-world data presents significant performance gains over state-of-the-art baselines in a variety of applications, including personality prediction, demographics inference, and user identification.
\end{abstract}




\maketitle


\input{intro}

\input{model}
\input{solution_org}

\input{eval}
\input{relate}
\input{conclusion}

\bibliographystyle{aaai}
\bibliography{refs}

\end{document}

%% file: intro.tex
\section{Introduction}
\label{sec:intro}

The wide proliferation of wearables and mobile devices is revolutionizing health and wellness with the potential of data and personalized insights at one's fingertips ~\cite{op2018fatigue}. These wearables generate chronologically ordered streams (\eg, the series of heart rate measurements) or point-in-time activity (\eg, the number of steps) or general summarization of the day (\eg, move goals). Collectively, these data provide an insight abut quantified self, and are also providing an unprecedented opportunity to infer qualitative attributes about an individual, including demographics, social network, opinions, beliefs, personality, and job performance. 

In this paper, we pose the problem of inferring personal attributes about an individual (qualified self) from the quantified data generated from wearables. Specifically, we focus on the challenges presented from the chronologically ordered data generated from the wearables (such as the heart rate data). This data is notated as  \emph{wearable-sensory time series} in this paper.
We posit that it is critical to effectively model these these wearable-sensory time series data to fully realize their benefit for a wide spectrum of applications, such as personality detection~\cite{cao2017deepmood}, job performance prediction, health and wellness state assessment and prediction, user identification~\cite{wang2018supervised} and demographics inference~\cite{wang2016your}. 

However, wearable-sensory time series data present a number of challenges, including temporal dependencies, incompleteness, intra-sensor / individual variability, and inter-individual variability. In this paper, we address these challenges by developing a generalized representation learning algorithm, \emph{\model}, and demonstrate how physiological responses as captured in the heart rate is predictive and indicative of several personal attributes, including demographics and personality. 

\vspace{-0.1in}
\paragraph{Limitations of Current Work.} The contemporary body research in time series has focused on wavelet-based frequency analysis~\cite{sun2015novel} and / or motif discovery methods~\cite{zhu2018matrix}.  \emph{First}, while we can extract discriminating and independent features using wavelet decomposition approaches, it still involves manual effort and domain-specific expert knowledge (\eg, medical knowledge)~\cite{bai2018interpretable}. \emph{Second}, discovering motifs are computationally expensive and require the repeated process of searching for optimal motifs from candidates~\cite{liu2015efficient}. 

Motivated by these limitations, recent research has also focused on representation learning on sensory data~\cite{ballinger2018deepheart,wang2018not,ni2019modeling}. But these works are also limited as they assume complete time series and fixed length sequences, thus not  sufficiently addressing the following challenges.  \emph{First}, the wearable-sensory time series data is of variable-length ranging from several days to months, since the active sensing time period may vary from person to person~\cite{liu2016human}. \emph{Second}, readings of sensors are usually missing / incomplete at different time periods for various reasons (\eg, sensor communication errors or power outages or compliance)~\cite{manashty2018concise}. \emph{Third}, the success of these supervised deep neural network models largely relies on substantial ground truth~\cite{choi2017gram}, which is often scarce. 


\vspace{-0.1in}
\paragraph{This paper. } In light of these challenges and limitations, we developed a time series representation learning model, \emph{\model}, that addresses the weaknesses of the aforementioned methods and learns generalized embeddings, leading to more accurate models to infer and predict user demographics, personality, work performance attributes, and health states. Specifically, \emph{\model} addresses the following {\bf{key challenges}}:   \\
\noindent \underline{1)} Handling variable-length time series and data incompleteness. A straightforward way to address variable-length input is to resize each input time series into a fixed-length vector. However, the resulting vector, derived by interpolation when input length differs from pre-defined length, has less control of variable time series resolutions and fails to capture the consistent time granularity for different individuals, leading to inferior representation. An alternative approach is sequence padding with specific symbols so that all time series are as long as the longest series in the dataset~\cite{srivastava2015training}. However, an associated challenge with sequence padding is that all time series data have been artificially created with fixed-length inefficiently~\cite{doetsch2017comprehensive}. Consequently, artificially created sequences with padding operation may not properly reflect their time-evolving distributions. This is especially critical when trying to draw insights from physiological response data such as heart rate. Thus, a representation learning method should be able to deal with variable length and incomplete data. \\
\noindent \underline{2)} Differentiating unique patterns from time series data with similar trends. In addition, the wearable-sensory data collected from different people may have very similar distributions~\cite{yang2006adaptive}. For example, for adults 18 and older, a normal resting heart rate (\ie number of heart beats per minute) is between 60 and 80 beats per minute~\cite{ballinger2018deepheart}.
However, there are localized patterns among individuals (intra-sensor / individual variability) that offer important nuances on similarities and differences. Thus, a representation learning method should be able to capture the global and local characteristics of such time series data to effectively compute similarities among individuals. 
To summarize, our {\bf{main contributions}} are as follows.  

We develop a novel  and effective representation learning model, \emph{\model}, that successfully addresses the aforementioned challenges to learn a generalizable  lower-dimensional and latent embedding from wearable sensory data. 
In \emph{\model}, we first segment the heart rate data collected from individuals into multiple day-long time series because human behavior to capture day-long regularities. Then, a deep autoencoder architecture is developed to map high-dimensional time-specific (\ie, daily) sensor data into the same latent space with a dual-stage gating mechanism. These learned embedding vectors are capable of not only preserving temporal variation patterns, but also largely alleviating the data missing issue of wearable-sensory time series. After that, we leverage a temporal pattern aggregation network to capture inter-dependencies across time-specific representations based on the developed position-aware multi-head attention mechanism. During the training process, a Siamese-triplet network optimization strategy is designed with the exploration of implicit intra-series temporal consistency and inter-series relations. 
We evaluate \emph{\model} on two real-world wearable-sensory time series data representing a diverse group of human subjects and two different sensing devices. Our experiments demonstrate that \emph{\model} outperforms the state-of-the-art methods in various applications, including user identification, personality prediction, demographics inference, and job performance prediction, thereby asserting the quality of the learned embedding. To the best of our knowledge, this is the most comprehensive study on addressing challenges of wearable time series data by developing a novel \emph{\model}, considering two real-world data representative of different human subjects and environments, and applying to a number of application scenarios.

%% file: model.tex
\section{Problem Formulation}
\label{sec:model}

We first introduce preliminary definitions and formalize the problem of wearable-sensory time series representation learning. We use bold capital and bold lower case letters to denote matrices and vectors, respectively.




\begin{Dfn}
\textbf{Wearable-Sensory Time Series}. Suppose there are $I$ users ($U=\{u_1,...,u_i,...,u_I\}$), and  $\textbf{X}_i=(x_i^1,...,x_i^j,...,x_i^{J_i})$ ($\textbf{X}_i \in \mathbb{R}^{J_i}$) to denote the series of temporally ordered wearable-sensory data (\eg, heart rate measurement --- the number of times a person's heart beats per minute) of length $J_i$ collected from user $u_i$. In particular, each element $x_i^j $ represents the $j$-th measured quantitative value from user $u_i$ ($1\leq j\leq J_i$ and $J_i$ may vary from user to user. Each measurement $x_i^j$ is associated with a timestamp information $t_i^j$ and thus we define a time vector $\textbf{T}_i=(t_i^1,...,t_i^j,...,t_i^{J_i})$ to record the timestamp information of sequential wearable sensor data $\textbf{X}_i$.
\end{Dfn}


Since the wearable-sensory data is collected over different time periods (\eg, with different start and end date), the sequence lengths usually vary among different time series~\cite{liu2016human}. Additionally, readings of wearable sensors are usually lost at various unexpected moments because of sensor or communication errors (\eg, power outages)~\cite{manashty2018concise}. Therefore, the wearable-sensory data often exhibit variable-length and missing values. \\\vspace{-0.1in}


\emph{Problem Statement}. \textbf{Wearable-Sensory Time Series Representation Learning}. Given a wearable-sensory time series $\textbf{X}_i$ with variable-length and missing values, the objective is to learn the $d$-dimensional latent vector $\textbf{Y}_i\in \mathbb{R}^d$ that is able to capture the unique temporal patterns of time series $\textbf{X}_i$.\\\vspace{-0.1in}

The output of the model is a low-dimensional vector $\textbf{Y}_i$ corresponding to the latent representation of each wearable-sensory time series $\textbf{X}_i$ from user $u_i$. Notice that, although different time series $\textbf{X}_i (i\in [1,...,I])$ can be of any length, their representations are mapped into the same latent space. {\it{This learned time series representations, generated by \model can benefit various health and wellness tasks without the requirement of substantial training instance pairs.}}






%% file: solution_org.tex
\section{Methodology: \emph{\model}}
\label{sec:solution_org}


\begin{figure}
    \centering
    \includegraphics[width=0.7\textwidth]{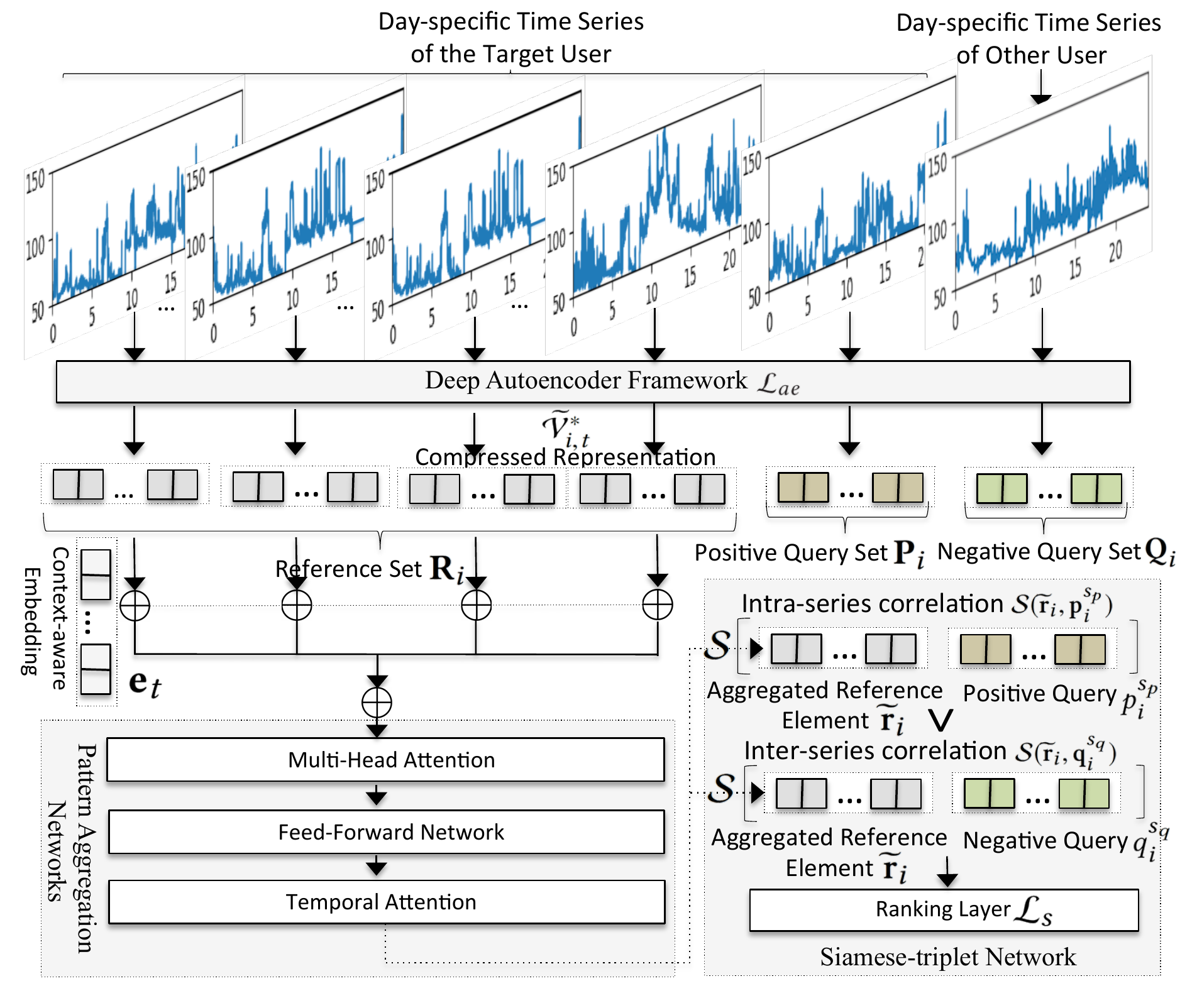}
    \caption{\model\ Framework. The training flow can be summarized as: (1) partition raw sensory data to a set of day-long time series; (2) map each day-long time series into a representation vector by Day-Specific Time Series Encoding Module (3) fuse day specific representations through Temporal Pattern Aggregation Network (4) update parameters based on reconstruction loss and Siamese-triple loss.}
    \label{fig:autoencoder}
    \vspace{-1em}
\end{figure}


\paragraph{Time Series Segmentation}
Considering that periodicity has been demonstrated as an important factor that governs human sensory data (\eg, heart rate) with time-dependent transition regularities~\cite{nakamura2018cross}, and the sensed time series data are often variable-length, we first partition the wearable-sensory time series (\ie, $\textbf{x}_i$) of each user $u_i$ into $T$ (indexed by $t$) separated day-long time series ($T$ may vary among users). 

\begin{Dfn}
\textbf{Day-long Time Series $\mathbf{x}_{i,t}$}. Each $t$-th divided day-long time series of $\textbf{x}_i$ is denoted as $\textbf{x}_{i,t} \in \mathbb{R}^{K}$, where $K=1440$ is the number of minutes included in one day. In $\textbf{x}_{i,t}$, each element $x_{i,t}^k$ is the measurement from user $u_i$ at the $k$-th time step in $\textbf{x}_{i,t}$. Due to the data incompleteness issue of the collected time series, we set the element $x_{i,t}^k$ as 0 to keep equally-spaced intervals for missing measurement.
\end{Dfn}

Leveraging the {\it{Day-long Time Series}}, \emph{\model} consists of three components: i) Day-specific time series encoding; ii) Temporal pattern aggregation network; and iii) Siamese-triplet network optimization strategy.

\subsection{Day-Specific Time Series Encoding}
To explore the underlying repeated local patterns and reduce dimensions of day-long time series data, we first propose a convolution autoencoder module to map each individual series $\textbf{x}_{i,t}$ into a common low-dimensional latent space. In general, the encoder first takes day-long time series as the input and then translates it into a latent representation which encodes the temporal pattern. Then, the decoder network reconstructs the data which is identical with the input $\textbf{x}_{i,t}$ in the ideal case. 
To keep the input dimension consistent with complete day-specific time series, we impute missing values with zero, as typically done. Since the imputed values are treated same as other valid inputs when applying kernels in each convolutional layer, the generated features can get incorrectly encoded and further lead to error propagation from lower to higher layers. To resolve this issue stemming from zero padding\hide{To mitigate the undesired effects from zero padding}, we propose a dual-stage gating mechanism into the convolution autoencoder module.
More specifically, each layer of our deep autoencoder framework is a four-step block: i) convolution network; ii) channel-wise gating mechanism; iii) temporal gating mechanism; and iv) pooling operation. Figure~\ref{fig:autoencoder} presents the architecture of our deep autonencoder module.

\subsubsection{\bf Convolutional Layers}
Firstly, we apply convolutional layers to encode the local pattern of day-long time series $\textbf{x}_{i,t}$. \hide{Specifically, we feed $\textbf{x}_{i,t}$ into a number of convolutional layers.} Let's denote $\mathbf{V}_{i,t}^{l-1}$
as the feature map representation of $(l-1)$-th layer.
The output of $l$-th layer is given as:
\begin{align}
\small
\mathbf{V}_{i,t}^l = f(\textbf{W}^l_c * \mathbf{V}_{i,t}^{l-1} + \textbf{b}^l_c)
\end{align}
\noindent where \hide{\begin{small}$\mathbf{V}_{i,t}^l=\mathbf{x}_{i,t}$\end{small}, }$f(\cdot)$ is the activation function and $*$ denotes the convolutional operation. $\textbf{W}^l_c$ and $\textbf{b}^l_c$ represents the transformation matrix and bias term in $l$-th layer, respectively.

\subsubsection{\bf Channel-Wise Gating Mechanism} 
The goal of the channel-wise gating mechanism is to re-weight hidden units by exploiting the cross-channel dependencies and selecting the most informative elements from the encoded feature representation $\mathbf{V}_{i,t}^l$~\cite{hu2018squeeze}.  To exploit the dependencies over channel dimension, we first apply temporal average pooling operation $\mathcal{F}_{pool}^{cg}(\cdot)$ on the feature representation $\mathbf{V}_{i,t}^l$ over temporal dimension ($1\!\!\leq\!\! k\!\!\leq\!\! K$) to produce the summary of each channel-wise representation as:
\begin{align}
\small
\textbf{z}_{i,t}^{l} = \mathcal{F}_{pool}^{cg}(\mathbf{V}_{i,t}^l) = \frac{1}{K} \sum_{k=1}^K \mathbf{V}_{i,t,k,:}^l
\label{equ:channel_1}
\end{align}
\noindent where $\textbf{z}_{i,t}^{l}$ is the intermediate representation of $\mathbf{V}_{i,t}^l$ after average pooling operation over temporal dimension. Then, our channel-wise gating mechanism recalibrates the information distribution among all elements across channels as:
\begin{align}
\small
\textbf{a}_{i,t}^{l} = \text{Sigmoid} (\textbf{W}_2^{cg} \cdot \text{ReLU}(\textbf{W}_1^{cg} \cdot \textbf{z}_{i,t}^{l}))
\label{equ:channel_2}
\end{align}
\noindent where $\textbf{a}_{i,t}^{l}$ denotes the channel-wise importance vector in which each entry is the each channel's importance. $\textbf{W}_1^{cg}$ and $\textbf{W}_2^{cg}$ are learned parameters of fully connected networks. Finally, the channel-wise representations $\widetilde{\mathbf{V}}_{i,t}^l$ is learned as:
\begin{align}
\small
\widetilde{\mathbf{V}}_{i,t,:,c}^l
= \mathbf{V}_{i,t,:,c}^l \textbf{a}_{i,t,c}^{l}
\label{equ:channel_3}
\end{align}
\noindent where $c$ is the channel index.

\subsubsection{\bf Temporal-wise Gating Mechanism}
To re-weight the hidden units by capturing the temporal dependencies of feature representation across time steps, we also apply a gating mechanism on temporal dimension to further learn focus points in the time-ordered internal feature representation $\widetilde{\mathbf{V}}_{i,t}^l$ (output from the channel-wise gating mechanism).
Similar to the gating mechanism procedures encoded in Equations~\ref{equ:channel_1}, \ref{equ:channel_2}, and \ref{equ:channel_3},
we first apply the channel average pooling operation on feature representation $\widetilde{\mathbf{V}}^l_{i,t}$ over channel dimension and get $\widehat{\mathbf{V}}_{i,t}^l
$ as the summarized feature representation which jointly preserve the channel-wise and temporal dependencies.



\subsubsection{\bf Encoder-Decoder Configuration.} We present the architecture configuration of our deep autoencoder module.

\noindent \textbf{Encoder. } Given the day-long time series $\textbf{x}_{i,t} \in \mathbb{R}^K$ 
, we use ReLU activation function with $5$ convolutional layers (\ie, Conv1-Conv5) followed by channel-wise, temporal-wise gating mechanism and pooling layers. 
Particularly, Conv1-Conv5 are configured with the one-dimensional kernel with $\{9,7,7,5,5\}$ and filter sizes with $\{32,64,64,128,128\}$, respectively. 
Then, we perform the flatten operation on the output to generate a one-dimensional feature representation and feed it into a fully connected layer with $\mathrm{Tanh}$ activation function to generate the final latent representation $\mathbf{g}_{i,t}$ corresponding to the time series $\textbf{x}_{i,t}$. Note that the number of layers, kernel sizes, and filter sizes are hyperparameters that may be configurable and could vary by specific sample interval(\eg, 30 seconds) of sensory data.\\

\noindent \textbf{Decoder. } The decoder is symmetric to the encoder in terms of the layer structure. First, the representation $\mathbf{g}_{i,t}$ is uncompressed by a single-layer with $\mathrm{Tanh}$ activation function and then followed by a series of five deconvolutional layers with ReLU activation function. The kernel and filter sizes is in reverse order to be symmetric to the encoder architecture configuration. Channel-wise and Temporal-wise are applied in the first four layer after deconvolutional operation.\\


\noindent \textbf{Loss Function in Deep Autoencoder Module}: We formally define the reconstruction loss function as follows:
\begin{align}
\small
\mathcal{L}_{ae}^{i} = \sum_{t}||\mathbf{m} \odot (\mathcal{D}(\mathcal{E}(\textbf{x}_{i,t})) - \textbf{x}_{i,t})||^2_2
\end{align}
\noindent where $\mathbf{m} \in \mathbb{R}^K$ is a binary mask vector corresponding to each element in $\mathbf{x}_{i,t}$. In particular, 
$m^k=1$ if $x_{i,t}^k\neq 0$(i.e., has measurement) and $m^k=0$ otherwise. $\odot$ is the element-wise product operation, $\textbf{x}_{i,t}$ is the input, $\mathcal{E}(\cdot)$ and $\mathcal{D}(\cdot)$ represents the encoder and decoder function.


\begin{figure}
    \centering
    \includegraphics[width=0.7\textwidth]{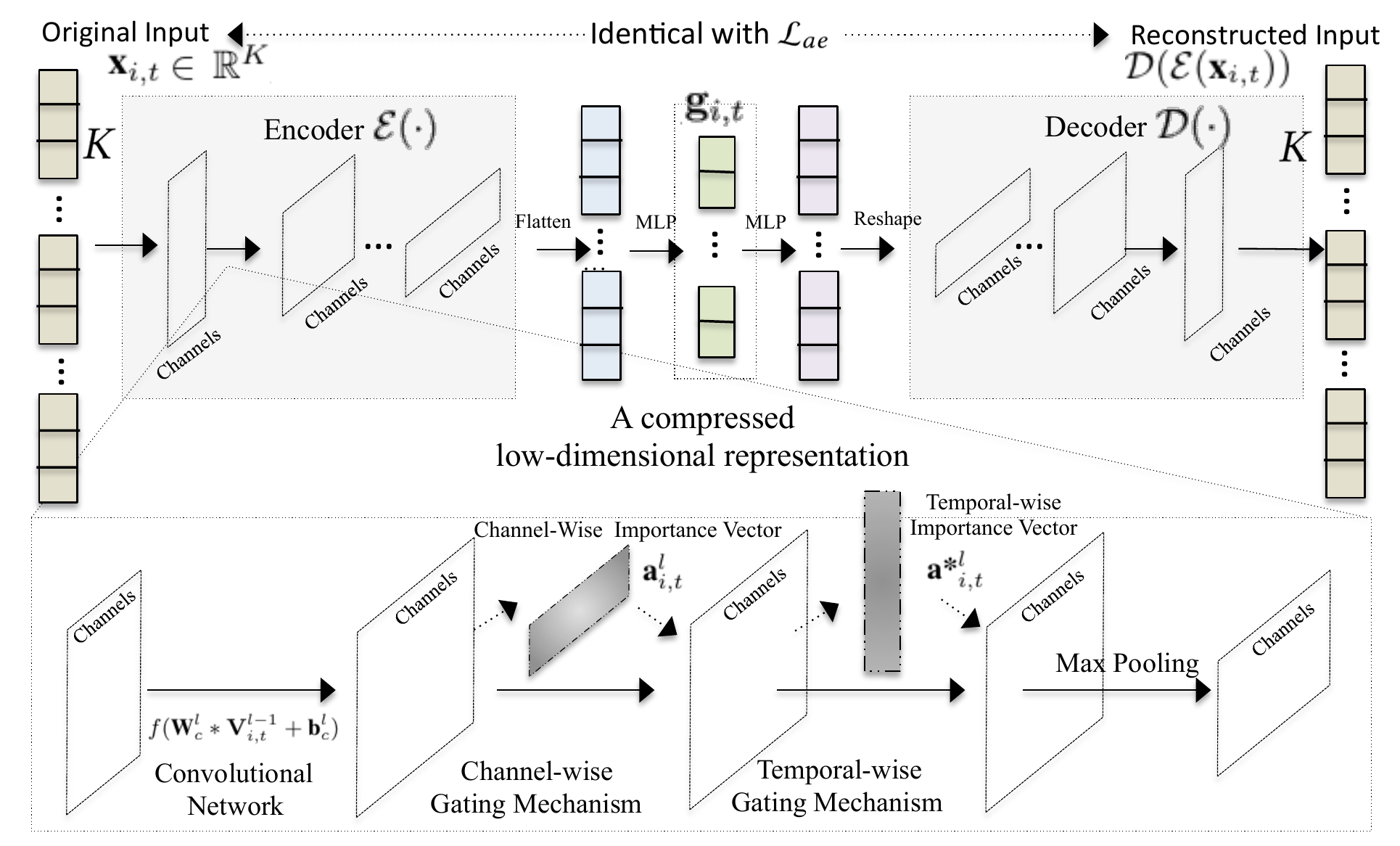}
    \caption{Illustration of Day-Specific Time Series Encoding.}
    \label{fig:autoencoder}
    \vspace{-1em}
\end{figure}

\subsection{Temporal Pattern Aggregation Network}
\begin{figure}
    \centering
    \includegraphics[width=0.7\textwidth]{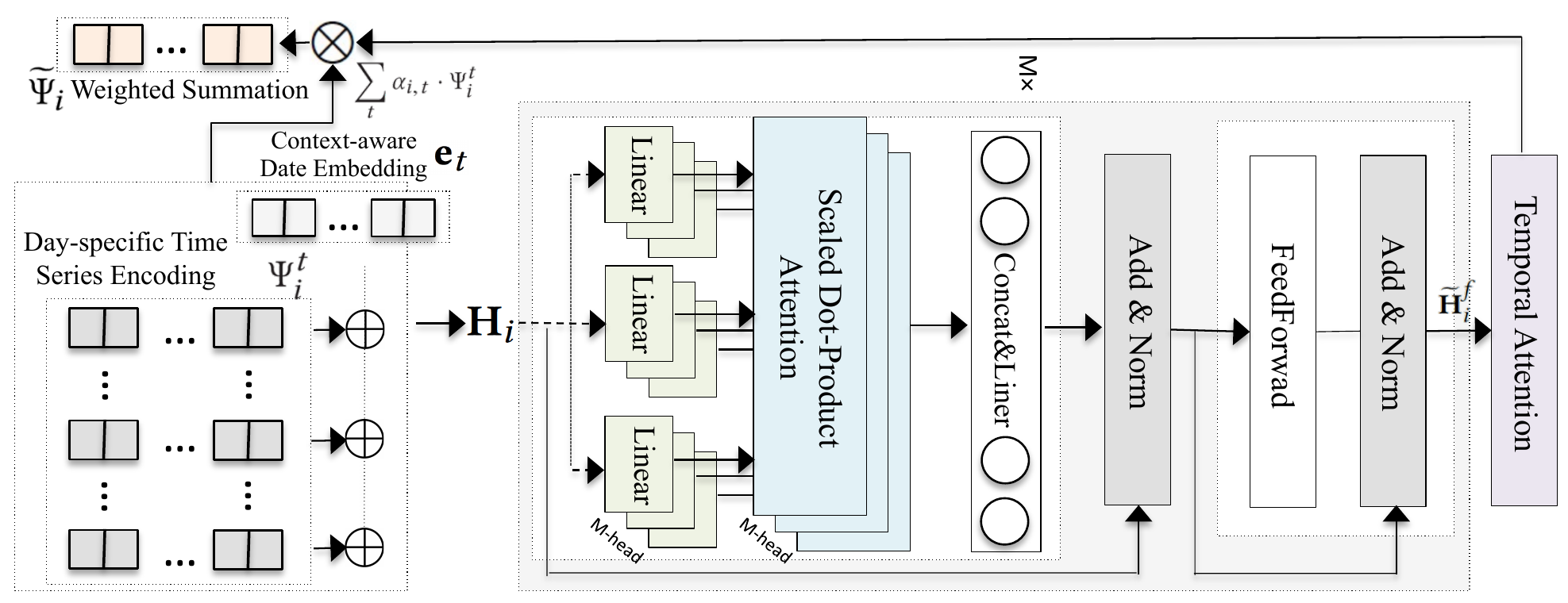}
    \caption{Temporal Pattern Aggregation Network.}
    \label{fig:fusion_process}
    \vspace{-1em}
\end{figure}

While applying the autoencoder framework to map day-long time series into a low-dimensional vector, the fusion of day-specific temporal patterns presents a challenge. To address this challenge\hide{ and aggregate the encoded temporal-wise patterns}, we develop a temporal pattern aggregation network which promotes the collaboration of different day-specific temporal units for conclusive cross-time representations. Figure~\ref{fig:fusion_process} shows the architecture of our temporal pattern aggregation network which consists of three major modules: i) context-aware time embedding module; ii) multi-head aggregation network; and iii) temporal attention network.



\subsubsection{\bf Context-Aware Time Embedding Module}
From the deep autoencoder module, given the time series $\mathbf{x}_i$ of user $u_i$, we learn a set of date-ordered latent representations with size of $T$, \ie, $\mathcal{G}_{i} = \{\mathbf{g}_{i,1},...,\mathbf{g}_{i,T}\}$.
In reality, different people may exhibit different wearable-sensory data distribution due to specific daily routines~\cite{probst2016emotional}. To incorporate the temporal contextual signals into our learning framework, we further augment our model with a time-aware embedding module, which utilizes the relative time difference between the last time step and each previous one. For example, given a time series with three date information $\{$2018-10-01, 2018-10-20, 2018-10-25$\}$, we generate a date duration vector as $\{24, 5, 0\}$ (the day duration between 2018-10-01 and 2018-10-25 is $24$ days). To avoid the occurrence of untrained long date duration in testing instances, we adopt timing signal method~\cite{vaswani2017attention} to represent a non-trainable date embedding. Formally, the vector $\mathbf{e}_t$ of $t$-th day is derived as:
\begin{align}
\small
e_{t, 2i} = \text{sin}(\frac{t}{10000^{2i/d_e}});~~e_{t, 2i+1} = \text{cos}(\frac{t}{10000^{2i/d_e}})
\end{align}
where \hide{$t$ is the relative time value and} $d_e$ is the embedding dimension ($2i+1$ and $2i$ are the odd and even index in the embedding vector). New context-aware latent vector $\mathbf{h}_{i,t} \in \mathbb{R}^{d_e}$ is generated by element-wise adding each day-specific feature representation $\mathbf{g}_{i,t}$ and date embedding $\textbf{e}_t$, to incorporate the temporal contextual signals into the learned embeddings.



\subsubsection{\bf Multi-Head Aggregation Network.}

During the pattern fusion process, we develop a multi-head attention mechanism that is integrated with a point-wise feedforward neural network layer to automatically learn the quantitative relevance in different representation subspaces across all context-aware temporal patterns. 
Specifically, given the $i$-th time series, we feed all context-aware day-specific embeddings $\textbf{H}_{i} = \{\textbf{h}_{i,0}, \cdots, \textbf{h}_{i,T}\}$ into a multi-head attention mechanism. Here, $M$-heads attention conducts the cross-time fusion process for $Q$ subspaces. Each $q$-th attention involves a separate self-attention learning among $\textbf{H}_{i}$ as:
\begin{align}
\small
\widetilde{\textbf{H}}_{i}^q = \text{softmax}(\frac{ \textbf{W}_1^{q} \cdot \textbf{H}_{i}(\textbf{W}_2^{q} \cdot \textbf{H}_{i})^T }{\sqrt{d_q}}) \textbf{W}_3^{q} \cdot \textbf{H}_{i} ,
\end{align}
where $\textbf{W}_1^{q}, \textbf{W}_2^{q}, \textbf{W}_3^{q} \in \mathbb{R}^{d_q\times d_e}$ represent the learned parameters of $q$-th head attention mechanism, and $d_q$ is the embedding dimension of $q$-th head attention,\ie, $d_q=d_e/Q$. Then, we concatentate each learned embedding vector $\widetilde{\textbf{H}}_{i}^q$ from each $q$-th head attention, and further capture the cross-head correlations as follows:
\begin{align}
\small
\widetilde{\textbf{H}}_{i} = \textbf{W}_{c} \cdot concat(\widetilde{\textbf{H}}_{i}^1,...,\widetilde{\textbf{H}}_{i}^Q)   
\end{align}
$\mathbf{W}_{c}\in \mathbb{R}^{d_e\times d_e}$ models the correlations among head-specific embeddings. Hence, we jointly embed multi-modal dependency units into the space with the fused $\widetilde{\textbf{H}}_{i}$ using the multi-head attention network. The advantage of our multi-head attention network lies in the exploration of feature modeling in different representation spaces~\cite{zhang2018gaan}. Then, we further feed the fused embedding $\widetilde{\textbf{H}}_{i}^c$ into two fully connected layers, which is defined as follows:
\begin{align}
\small
\widetilde{\textbf{H}}_{i}^f =\textbf{W}_2^f \cdot \text{ReLU}(\textbf{W}_1^f \cdot \widetilde{\textbf{H}}_{i} +\textbf{b}_1^f)+\textbf{b}_2^f,
\end{align}
where $\textbf{W}_1^f$, $\textbf{W}_2^f$ and $\textbf{b}_1^f$, $\textbf{b}_2^f$ are the weight matrix and bias in the feed-forward layer. In our \model\ framework, we perform multi-head attention mechanism twice.

\subsubsection{\bf Temporal Attention Network.}
To further summarize the temporal relevance, we develop a temporal attention network to learn importance weights across time. Formally, our temporal attention module can be represented as follows:
\begin{align}
\label{equ:query_attention_weight}
\small
\boldsymbol{\alpha}_i = \text{softmax}(\textbf{c}\cdot \text{Tanh}(\textbf{W}^{a} \widetilde{\textbf{H}}_{i}^f +\textbf{b}^{a}));\widetilde{\mathbf{g}}_{i} = \sum_{t} \alpha_{i,t}\mathbf{g}_{i,t}
\end{align}
Output $\widetilde{\textbf{H}}_i^f$ is first fed into a one fully connected layer and then together with the context vector $\textbf{c}$, it generates the importance weights $\boldsymbol{\alpha}_i$ through the softmax function. The aggregated embedding $\widetilde{\mathbf{g}}_{i}$ is calculated as a weighted sum of day-specific embeddings based on these learned importance weights. For simplicity, we denote our temporal pattern aggregation network as $\widetilde{\mathbf{g}}_{i}=\mathcal{A}(\mathcal{G}_{i})$.

\subsection{Siamese-triplet Network Optimization Strategy}

\hide{Our goal is to embed each individual wearable-sensory time series into low-dimensional spaces, in which every time series is represented as an embedding vector. We develop a Siamese-triplet network optimization strategy to jointly model the structural information of intra-series temporal consistency and inter-series correlations in our learning process. In particular, within a series of sensing data points (\eg, heart rate records), even though the measurements change over time, they do not change drastically within a short time period, since they belong to the same user~\cite{ukil2016heart}. Additionally, the heart rate measurements sampled from consecutive time intervals (\eg, days) of the same people may a more similar distribution as compared to sampling from different people~\cite{bhattacharjee2018heart}. Motivated by these observations, t}

Our Siamese-triplet network optimization framework aims to enhance the encoded user feature representations with the consideration of following constraint, \ie, making representation pairs from same user \hide{intra-series}closer to each other and representation pairs from different user \hide{inter-series data point pairs }further apart. We first define the following terms to be used in our optimization strategy.
\begin{Dfn}
\textbf{Reference Set $\mathcal{R}_i$}: we define $\mathcal{R}_i$ to represent the sampled reference set of user $u_i$. In particular, $\mathcal{R}_i = \{\textbf{r}_i^1,\cdots,\textbf{r}_i^{N^r}\}$ , where $N^r$ is size of reference set corresponding to $N^r$ sampled day-specific time series of $\textbf{x}_i$ defined in Definition 2. Each entry in $\mathcal{R}_i$ represents the $n_r$-th sampled time series from user $i$.
\end{Dfn}

\begin{Dfn}
\textbf{Positive Query Set $\mathcal{P}_i$}: we define $\mathcal{P}_i = \{\textbf{p}_i^1,\cdots,\textbf{p}_i^{N^p}\} $ to denote the positive query set of user with size of $N^p$. Specifically, every entry $\textbf{p}_i^{n^p}$ represents the $n^p$-th sampled day-specific time series from user $u_i$.
\end{Dfn}

\begin{Dfn}
\textbf{Negative Query Set $\mathcal{Q}_i$}: $\mathcal{Q}_i= \{\textbf{q}_i^1,\cdots,\textbf{q}_i^{N^q}\}$ is defined to as the negative query set for user $u_i$, which is the sampled $N^q$ day-specific time series from other users except user $u_{i}$, \ie, $u_{i'} (i'\neq i)$.
\end{Dfn}

Based on the above definitions, given a specific user $u_i$, 
we first aggregate the elements from user $u_i$'s reference set $\mathcal{R}_i$ as: $\widetilde{\mathbf{r}}_i=\mathcal{A}(\mathcal{R}_i)$, where $\mathcal{A}(\cdot)$ is the aggregation function which represents the developed temporal pattern aggregation network. Then, we compute the cosine similarity of aggregated reference element $\widetilde{\mathbf{r}}_i$ and each query (\ie, $\textbf{p}_i^{n^p}$ and $\textbf{q}_i^{n^q}$) from the generated positive query set $\mathcal{P}_i$ and negative query set $\mathcal{Q}_i$. Formally, the similarity estimation function $sim$ is presented as follows:
\begin{align}
\small
sim(\widetilde{\mathbf{r}}_i, \mathbf{p}_i^{n^p}) = \widetilde{\mathbf{r}}_i\cdot (\mathcal{E}(\textbf{p}_i^{n^p}))^T; ~~n^p\in [1,...,N^p]; \nonumber\\
sim(\widetilde{\mathbf{r}}_i, \mathbf{q}_i^{n^q}) = \widetilde{\mathbf{r}}_i\cdot (\mathcal{E}(\textbf{q}_i^{n^q}))^T; ~~n^q\in [1,...,N^q];
\end{align}
\noindent where $sim(\widetilde{\mathbf{r}}_i, \mathbf{p}_i^{n^p}) \in \mathbb{R}^1$ and $sim(\widetilde{\mathbf{r}}_i, \mathbf{q}_i^{n^q}) \in \mathbb{R}^1$. By capturing the temporal consistency of each individual user $u_i$ and inconsistency among different users, we optimize representations that preserve inherent relationships between each user's reference set and query set, \ie, time series embeddings from the same user are closer to each other, while embeddings from different users are more differentiated from each other. \hide{More specifically, the similarity between the aggregated reference element $\widetilde{\mathbf{r}}_i$ and the positive query $\textbf{p}_i^{s_p}$ should be larger, while the similarity between $\widetilde{\mathbf{r}}_i$ and the negative query $\textbf{q}_i^{s_q}$ should be smaller, \ie, $\mathcal{S}(\widetilde{\mathbf{r}}_i,\textbf{p}_i^{s_p}) > \mathcal{S}(\widetilde{\mathbf{R}}_i, \textbf{q}_i^{s_q})$.} Therefore, we formally define our loss function as follows:
\begin{small}
\begin{align}
\mathcal{L}_{s}^i=  \sum_{n^p}\sum_{n^q}\mathrm{max}(0, sim(\widetilde{\mathbf{r}}_i, \textbf{q}_i^{s_q})-sim(\widetilde{\mathbf{r}}_i, \textbf{p}_i^{s_p})+z)
\end{align}
\end{small}
\noindent where $z$ is the margin between two similarities. The objective function of joint model is defined as:
$\mathcal{L}_{joint} = \sum_i\mathcal{L}_{ae}^i + \lambda \mathcal{L}_{s}^i$, where $\lambda$ is the coefficient which control the weight of Siamese-triplet loss. The model parameters can be derived by minimizing the loss function. We use Adam optimizer to learn the parameters of \model. The model optimization process is summarized in Algorithm~\ref{alg:model}. 

\begin{algorithm}
    \SetKwData{User}{$\mathcal{U}$}
    \SetKwData{Item}{$\mathbf{I}$}
    \SetKwData{OneUser}{u}
    \SetKwData{OneItem}{i}
    
    \SetKwData{Train}{$\mathbf{X}$}
    \SetKwData{TrainBatch}{$\mathsf{T_{batch}}$}
    \SetKwData{TrainBatch}{\textsf{QueryBatch}}
    
    \SetKwData{Rtrue}{${r}_{t,a_k}$}
    \SetKwData{Rpred}{$\hat{r}_{t,a_k}$}
    
    \SetKwData{Step}{$s$}
    \SetKwData{Batch}{$b_{\textsf{size}}$}
    \SetKwData{Support}{$S_{\textsf{size}}$}
    \SetKwData{Positive}{$P_{\textsf{size}}$}
    \SetKwData{Negative}{$N_{\textsf{size}}$}
    \SetKwData{classPerEpisode}{$N_c$}
    \SetKwFunction{Sample}{sample}
    
    \SetKwFunction{LSTM}{LSTM}
    \SetKwFunction{RandomWalk}{RandomWalk}
    \SetKwData{Null}{\textsf{NULL}}
    \SetKwData{LSTMRes}{c}
    \SetKwFunction{MLP}{MLP}
    \SetKwData{OtherPara}{$\theta$}
    \SetKwFunction{Projection}{Projection}
    
    \SetKwData{Loss}{$\mathcal{L}$}
    
    \SetKwInOut{Parameter}{Paras}

    \KwIn{User set $\mathcal{U}$; reference set size $N^{r}$, positive query size $N^{p}$, negative query size $N^{q}$}
    \emph{Initialize all parameters}\;
    \While{not converge}
    {
        sample a set of users from user set $\mathcal{U}$\; 
        \ForEach{user $u_i$}
        {
            sample support set $\mathcal{R}_i$ and positive set $\mathcal{P}_i$\;
            sample negative set $\mathcal{Q}_i$\;
            feed each entry in $\mathcal{R}_i$, $\mathcal{P}_i$ and $\mathcal{Q}_i$ into autoencoder to get daily representations and compute $\mathcal{L}_{ae}^i$\; 
            aggregate support set $\mathcal{R}_i$ and derive $\mathcal{L}_{s}^i$\; 
        }
        \emph{update all parameters w.r.t $\mathcal{L}_{joint}=\sum_{i}\mathcal{L}_{ae}^i+\mathcal{L}_{s}^i$}\;
    }
    \caption{The Model Inference of \model.}
    \label{alg:model}
\end{algorithm}

\noindent \textbf{Unsupervised and semi-supervised learning scenarios}. \model\ is a general representation learning model that is flexible for both unsupervised (without labeled instances) and semi-supervised (limited number of labeled instances) learning. In semi-supervised learning, given the labeled time series and its target value, we take the learned representation vector as the input of a single-layer perceptrons with a combined loss function, \ie integrate the joint objective function $\mathcal{L}_{joint}$ with the loss function based on cross-entropy (categorical values) or MSE (quantitative values).

%% file: eval.tex
\section{Evaluation}
\label{sec:eval}
We comprehensively evaluate \emph{\model} on several inference and prediction tasks: user identification, personality prediction, demographic inference and job performance prediction. Our longitudinal real-world data comes from two different studies leveraging two different sensors and different population groups, thus allowing us to carefully vet and validate our findings. To robustly evaluate the accuracy and generalization of embeddings learned by \emph{\model}, within the context of the aforementioned prediction tasks, we aim to answer the following questions:


\begin{itemize}[leftmargin=*]
\item \textbf{Q1}: How does \emph{\model} perform compared with state-of-the-art representation learning methods for the wearable-sensory time series, as represented by heart rate?
\item \textbf{Q2}: How is the performance of \emph{\model} with respect to different training/testing time periods in user identification?
\item \textbf{Q3}: How do different components of \emph{\model} (\ie, deep autoencoder module, multi-head aggregation network and Siamese-triplet network optimization strategy) contribute to the \emph{\model} performance?
\item \textbf{Q4}: How do the key hyperparameters (\eg, reference set size $S_r$ and embedding dimension $d_e$) affect \emph{\model} performance?
\item \textbf{Q5}: How does  \emph{\model} lend itself for visualizing similarities and differences among individuals (interpretation)?
\end{itemize}


\subsection{Experimental Settings}

\subsubsection{\bf Data Description.}
We considered the heart rate time series data from two research projects~\footnote{Project names are anonymized for double-bind review}. For the purpose of the paper, we shall notate them based on the sensor: Garmin and Fitbit, respectively.\\\vspace{-0.1in}

\noindent \textbf{Garmin Heart Rate Data}.
This dataset comes from an on-going research study of workplace performance which measures the physiological states of employees in multiple companies. This dataset is collected from 578 participants (age between 21 to 68) by Garmin band from March 2017 to August 2018. Each measurement is formatted as (user id, heart rate, timestamp).\\\vspace{-0.1in}

\noindent \textbf{Fitbit Heart Rate Data}.
We collect this dataset from a research project at a University which aims to collect survey and wearable data from an initial cohort of 698 students (age between 17 to 20) who enrolled in the Fall semester of 2015. This dataset is collected by Fitbit Charge during the 2015/2016 and 2016/2017 academic years. 

\subsubsection{\bf Data Distribution.}
\begin{figure}[t!]
    \centering
    \subfigure[Garmin Data]{
        \centering
        \includegraphics[width=0.23\textwidth]{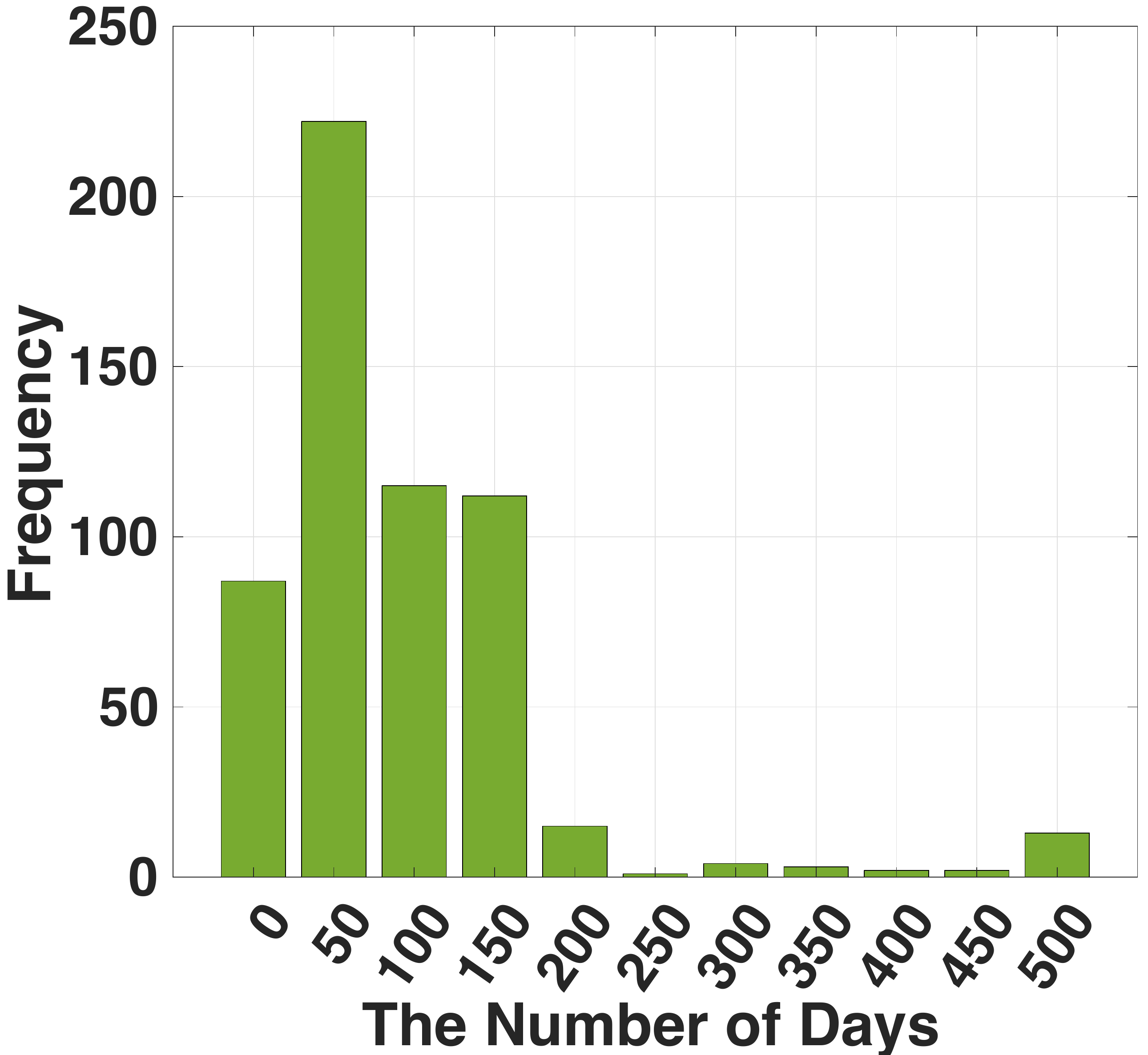}
        \label{fig:Macro}
        }
    \subfigure[Fitbit Data]{
        \centering
        \includegraphics[width=0.23\textwidth]{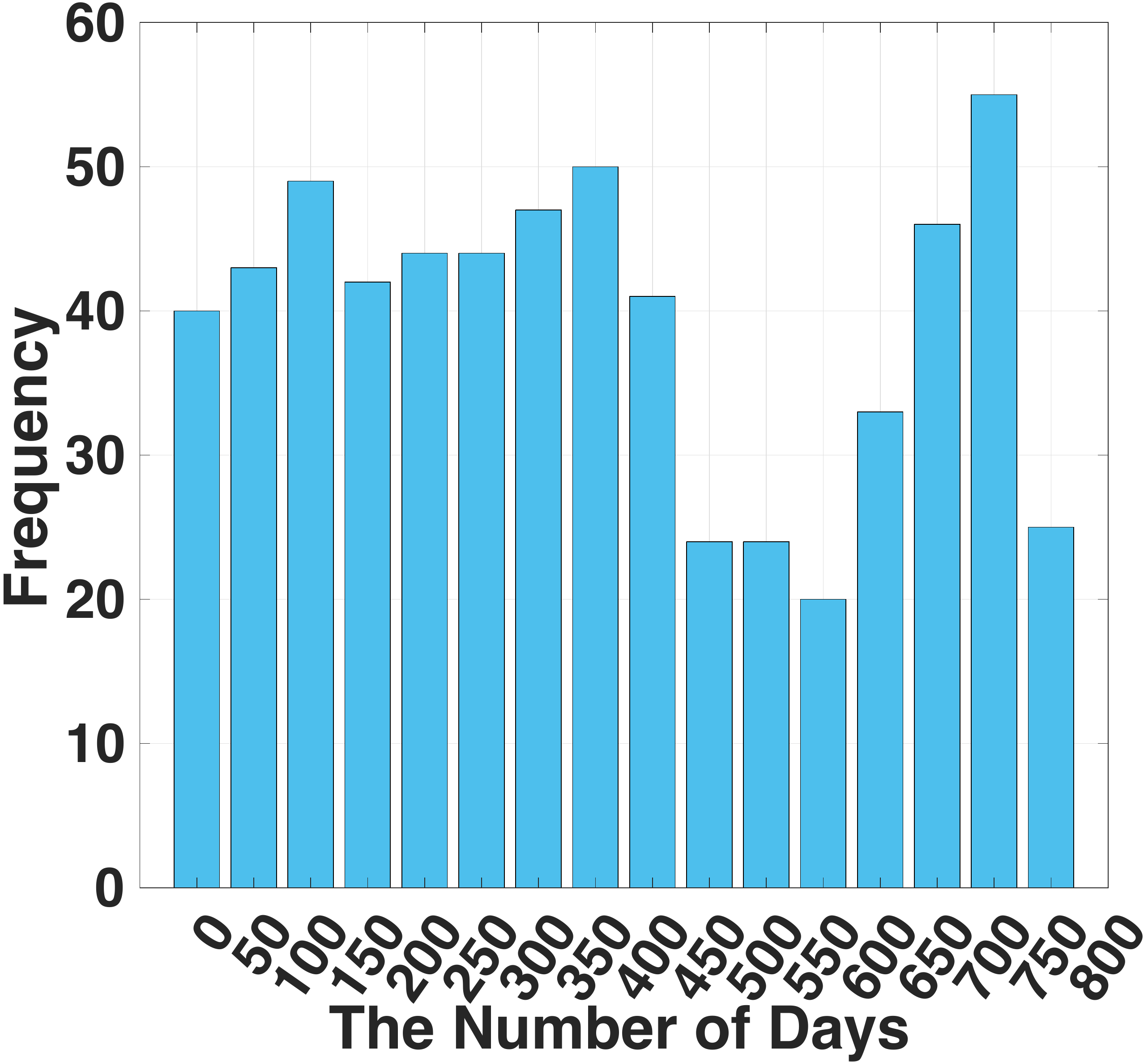}
        \label{fig:Micro}
        }
    \subfigure[Garmin Data]{
        \centering
        \includegraphics[width=0.23\textwidth]{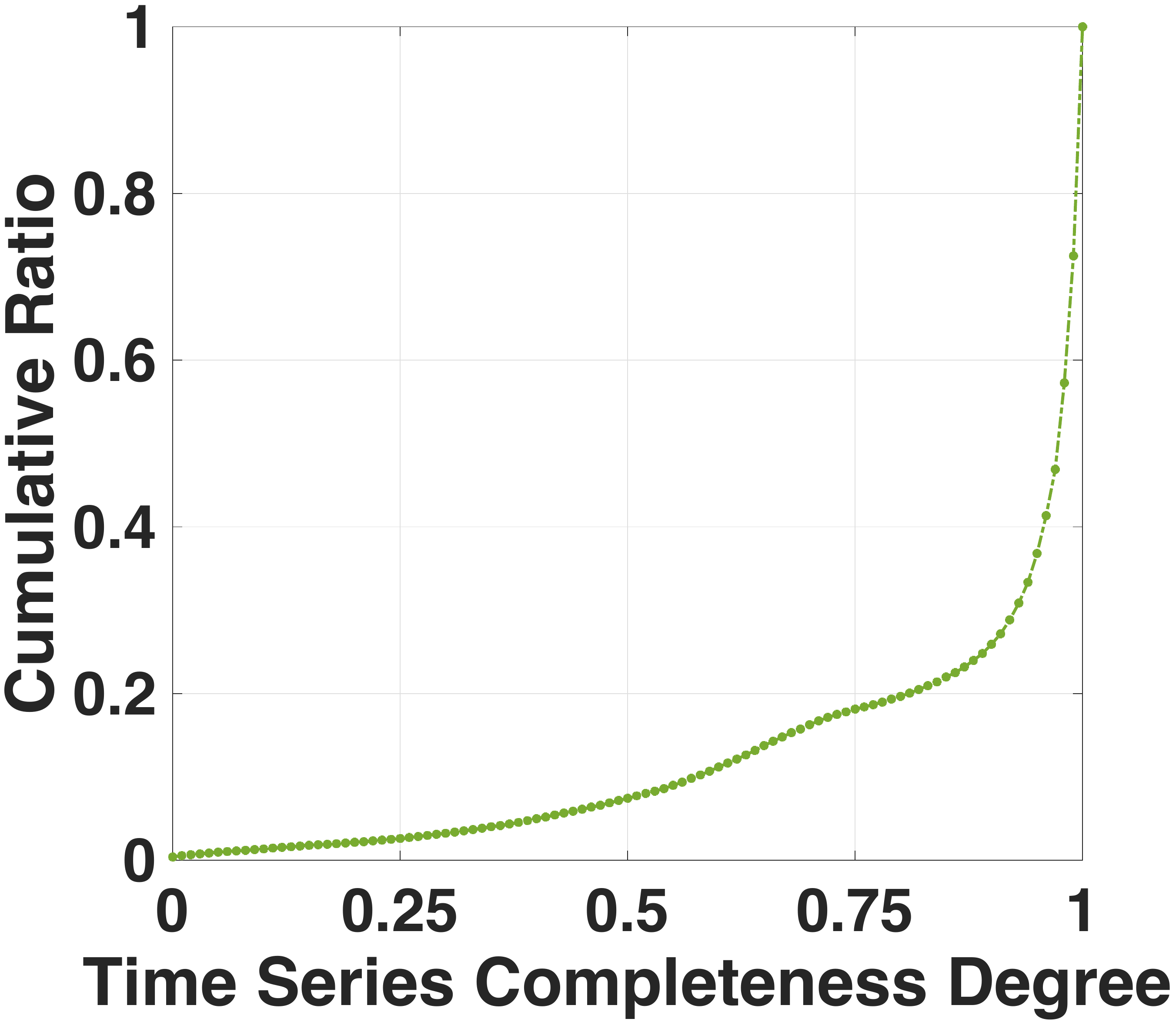}
        \label{fig:teserrea}
        }
    \subfigure[Fitbit Data]{
        \centering
        \includegraphics[width=0.23\textwidth]{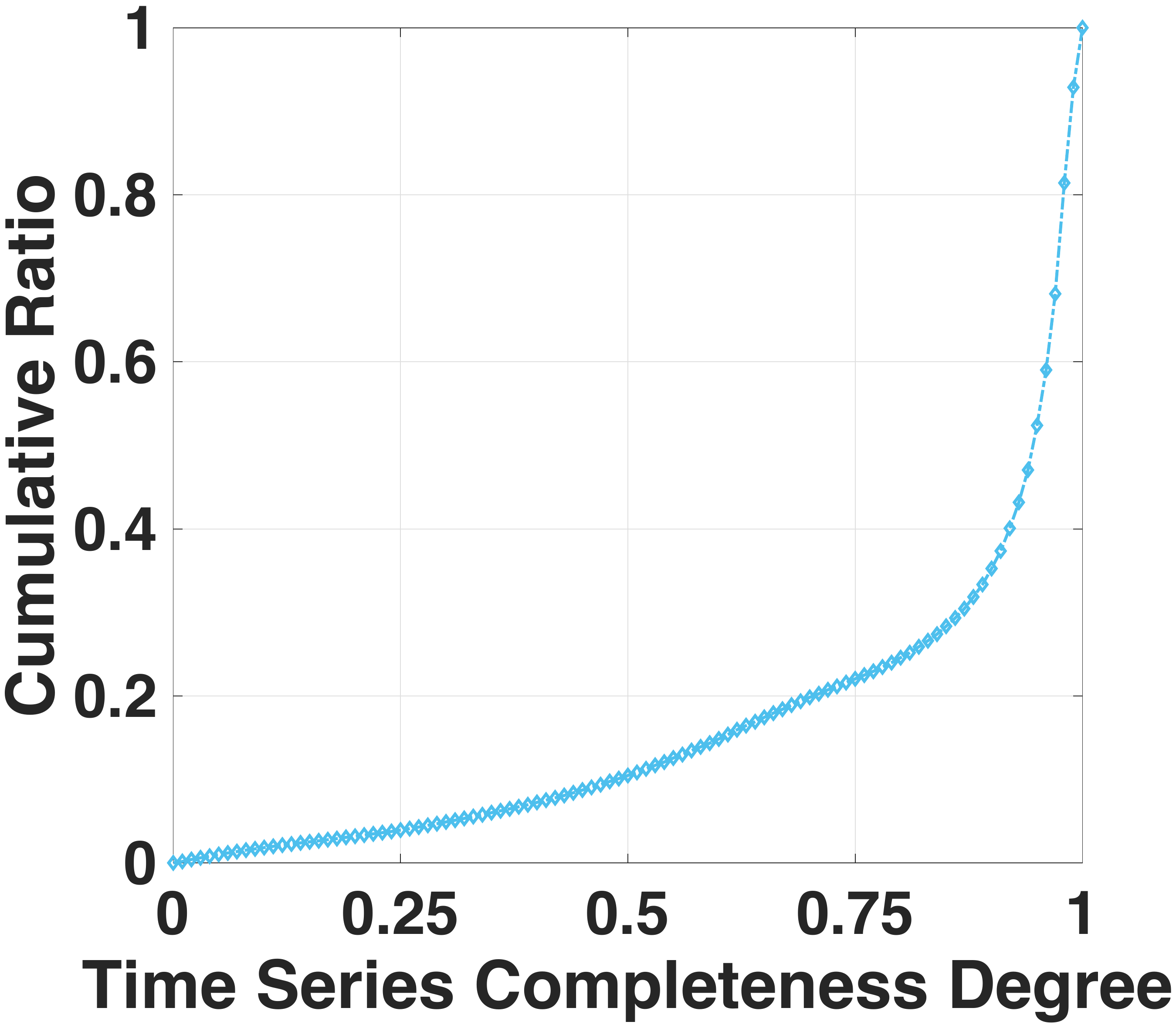}
        \label{fig:nethealth}
        }
    \vspace{-0.5em}
    \caption{Distribution of Time Series Length (a-b) and Completeness Degree (c-d).}
    \label{fig:dis_time_series}
\end{figure}

 Figure~\ref{fig:dis_time_series} shows the distribution of wearable-sensory time series in terms of time series length $J_i$ and completeness degree (\ie, the ratio of non-zero elements in day-specific time series $\textbf{x}_i^t$) on both Garmin and Fitbit heart rate data. As depicted in Figure~\ref{fig:dis_time_series} (a) and (b), different datasets have different time series distributions. Furthermore, Figure~\ref{fig:dis_time_series} (c) and (d) reveals that data incompleteness is ubiquitous, \eg, there exists more than 20\% day-specific time series with data incompleteness $<0.8$, which poses a further challenge for \emph{\model}. These observations are the main motivation to develop a temporal pattern aggregation network and a dual-stage gating mechanism for handling variable-length time series with incomplete data.


\begin{figure}[t!]
    \centering
    \subfigure[][User Demographic]{
        \centering
        \includegraphics[width=0.23\textwidth]{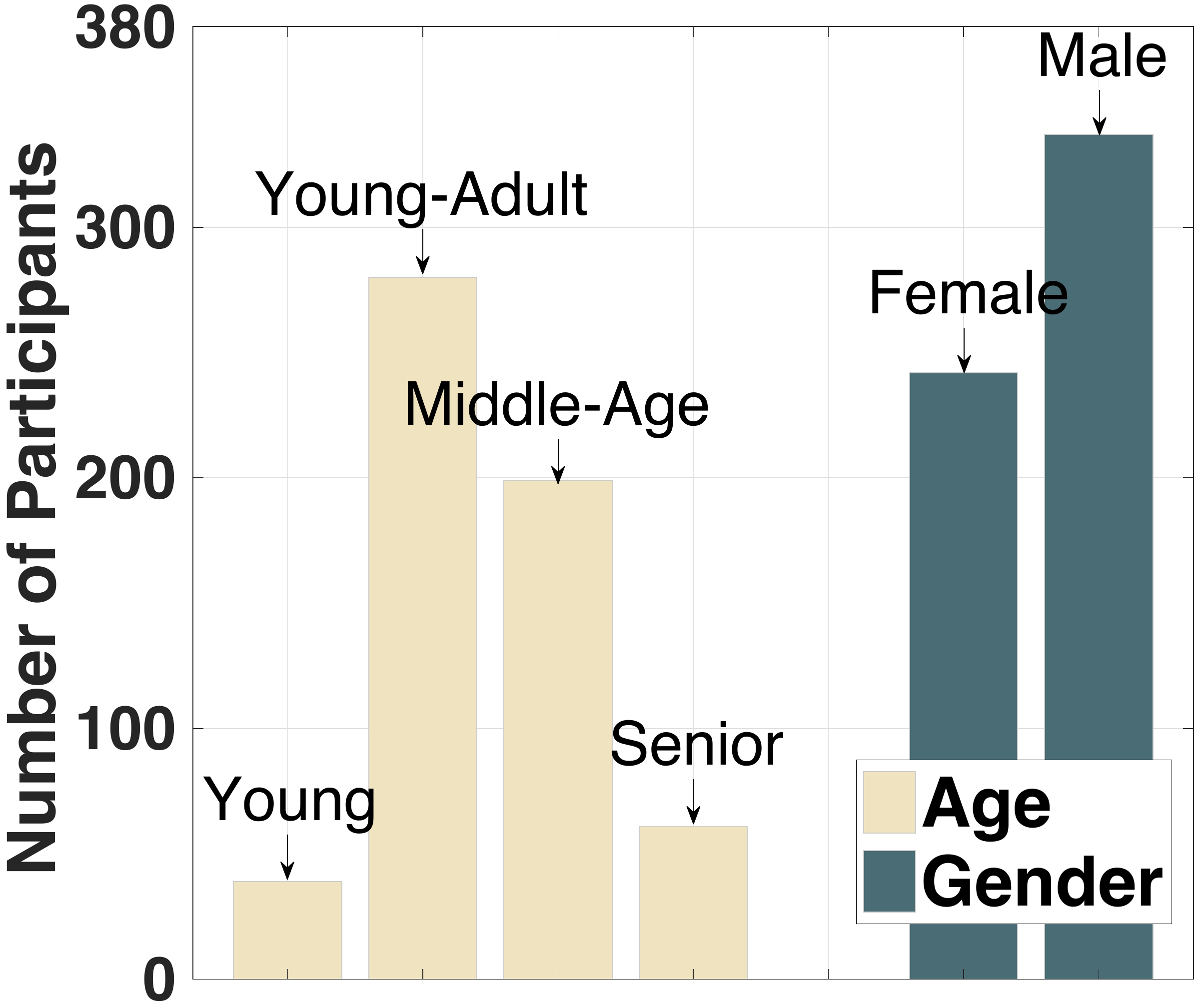}
        \label{fig:demongraphic}
        }
    \subfigure[][User Personality]{
        \centering
        \includegraphics[width=0.23\textwidth]{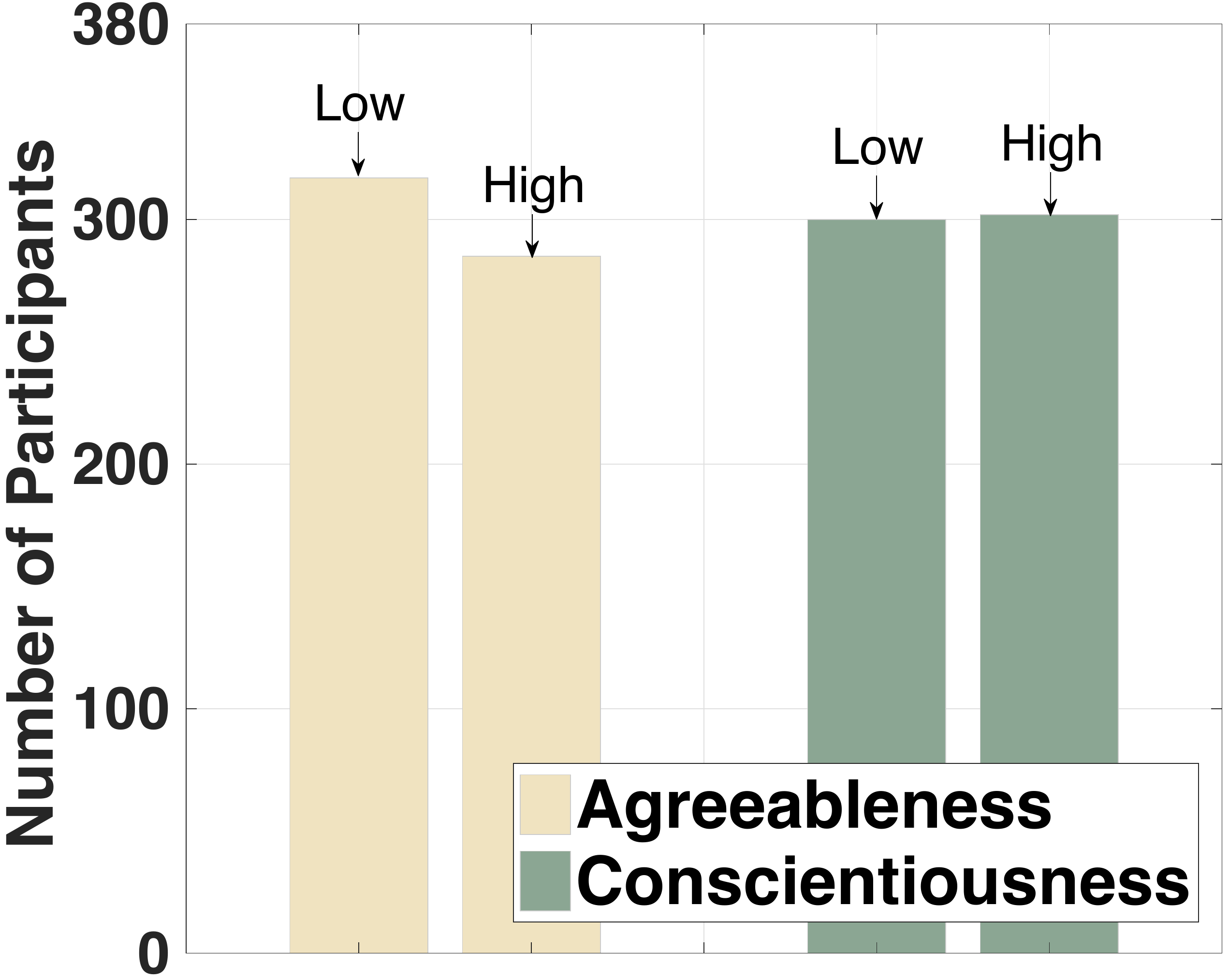}
        \label{fig:personality}
        }
    \subfigure[][Work Performance]{
        \centering
        \includegraphics[width=0.23\textwidth]{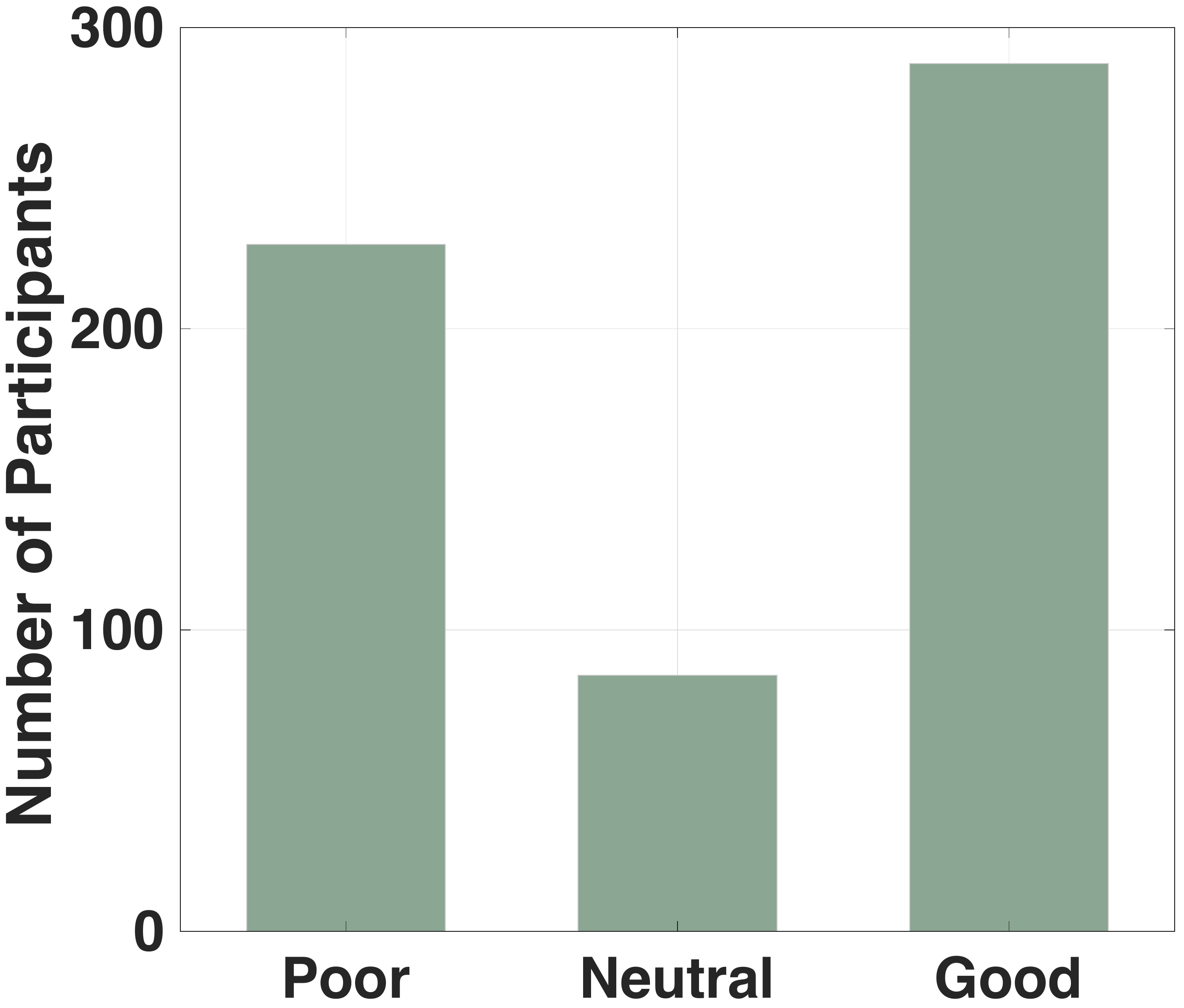}
        \label{fig:personality}
        }
    \subfigure[][Sleep Duration]{
        \centering
        \includegraphics[width=0.23\textwidth]{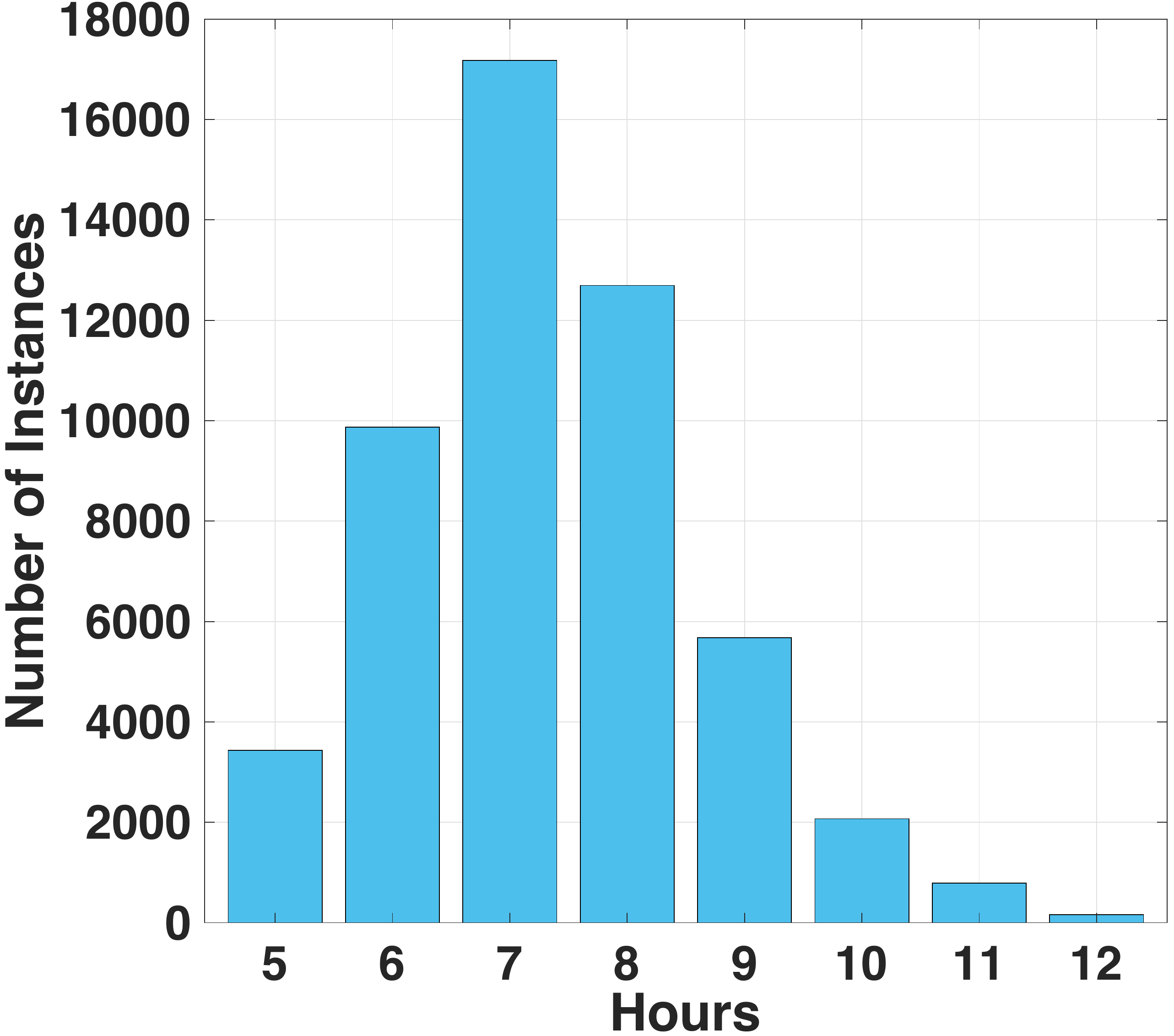}
        \label{fig:sleep}
        }
    \vspace{-0.5em}
    \caption{Data Statistics \wrt\ User Demographics, Personality and Sleep Duration.}
    \label{fig:data_statistics}
\end{figure}


\subsubsection{\bf Methods for Comparison.}
To justify the effectiveness of \emph{\model} for representation learning on wearable-sensory data, we compared it with the following contemporary representation learning methods:

\begin{itemize}[leftmargin=*]



\item \textbf{Convolutional Autoencoder (CAE)}~\cite{bascol2016unsupervised}: CAE is a representation learning framework by applying convolutional autoencoder to map the time series patterns into latent embeddings. \\\vspace{-0.1in}

\item \textbf{Deep Sequence Representation (DSR)}~\cite{amiriparian2018deep}: DSR is a general-purpose encoder-decoder feature representation model on sequential data with two deep LSTMs, one to map input sequence to vector space and another to map vector to the output sequence. \\\vspace{-0.1in}

\item \textbf{Multi-Level Recurrent Neural Networks (MLR)}~\cite{cheng2017recurrent}: MLR is a multi-level feature learning model to extract low- and mid-level features from raw time series data. RNNs with bidirectional-LSTM architectures are employed to learn temporal patterns.

\item \textbf{Sequence Transformer Networks (STN)}~\cite{oh2018learning}: STN is an end-to-end trainable method for learning structural information of clinical time-series data, to capture temporal and magnitude invariances. \\\vspace{-0.1in}

\item \textbf{Wave2Vec}~\cite{yuan2019wave2vec}: It is a sequence representation learning framework using skip-gram model by considering sequential contextual signals. Specifically, it takes the one-by-one data points that surround the target point within a defined window, to feed into a neural network for appear probability prediction. The window size and number of negative samples is set as 1 and 4, respectively.\\\vspace{-0.1in}

\item \textbf{DeepHeart}~\cite{bhattacharjee2018heart}: DeepHeart is a deep learning approach which models the temporal pattern of heart rate time-series data with the integration of the convolutional and recurrent neural network architecture. \\\vspace{-0.1in}


\end{itemize}
The initial parameter settings for each of the above methods are consistent with their original papers. 

\vspace{-0.1in}
\subsubsection{\bf Evaluation Protocols.}
To measure the effectiveness of \emph{\model} for representation learning on heart rate data in downstream applications, we conduct multiple tasks, including user identification, personality prediction, user demographic inference, job performance prediction and sleep duration prediction. These tasks reflect inferring qualified self from quantified self.  We summarize the details for five different tasks as follows:

\begin{itemize}[leftmargin=*]
\item \textbf{User Identification}: In the user identification evaluation, each of the aforementioned methods learns a mapping function to encode each of the day-specific time series data into a low-dimensional representation vector.  Then, given multiple day-specific time series data from one user and an unknown day-specific time series, the task is to predict whether the unknown time series is collected from the same user of the multiple day-specific time series. Specifically, we first map day-specific time series to embedding vector by utilizing each of the aforementioned methods. 
Then we aggregate embedding vectors of multiple day-specific time series to generate the reference vector. We apply mean-pooling operation on baselines and temporal aggregation network on \model\ during the aggregation process.
After that, we take the element-wise product between the reference vector and the embedding vector of the single day-specific time series as the input. Here, we adopt a logistic regression classifier to learn the model from the training data and generate the prediction. 
The True Positives and True Negatives are the heart rate series that are correctly identified for user $u_i$ or not, respectively, by the classifier. The False Positives and False Negatives are the heart rate series that are misclassified as belonging to user $u_i$ or not, respectively.


\item \textbf{Personality Prediction}: We considered two personality attributes of the participants, namely Conscientiousness and Agreeableness. We considered a classification task of a binary prediction on these attributes --- high or low for each participant $u_i$~\cite{mccrae1992introduction}.
In this task, the goal was to evaluate whether the embeddings learned from the heart rate are effective predictors of the personality types. We use the Logistic Regression classifier to make the prediction with the embedding as the feature space. The True Positives and True Negatives are the participants that are correctly classified by the classification method as high- and low-level, respectively. The False Positives and False Negatives are the high- and low-level participants that are misclassified, respectively. We only use the Garmin data for this task, as we only survey personality attributes for the subjects in the pool.


\item \textbf{User Demographic Inference}: We evaluate whether user demographics, such as gender and age are predictable by using a similar experimental construct as in~\cite{dong2014inferring}. The age information is categorized into four categories (\ie, Young: from 18 to 24, Young-Adult: from 23 to 34, Middle-age: from 39 to 49, senior: from 49 to 100).

\item \textbf{Job Performance Prediction}: We also evaluate the performance of predicting each participant's job performance (Individual Task Performance), which is categorized into three types, \ie, good, neutral, poor~\cite{conway1999distinguishing}. We only use the Garmin data for this task, as we only collect job performance information for the subjects in the Garmin pool. 

\item \textbf{Sleep Duration}: We further conduct the sleep duration estimation task to evaluate the quality of learned representations. We utilize the Linear Regression model which takes the embedding vectors of the target day-long time series as the input and predicts the sleep duration of that day as the output.  
\end{itemize}

We evaluate user identification and personality prediction tasks in terms of \emph{F1-score}, \emph{Accuracy} and \emph{AUC}. In addition, the user demographic inference and job performance prediction tasks, given multi-class scenario, are evaluated using \emph{Macro-F1} and \emph{Micro-F1}. The sleep duration task is evaluated in terms of \emph{MSE} and \emph{MAE}.

\subsubsection{\bf Training/Test Data Split.} We summarize the details of training/test data partition for different evaluation tasks as follows:
\begin{itemize}[leftmargin=*]

\item For demographic inference, personality prediction and job performance prediction task, we use the entire heart rate data for learning time series embeddings, and split the labels with $60\%$, $10\%$ and $30\%$ for training, validation and test, respectively.

\item For the user identification task, we
split the datasets in chronological order. We first use Garmin data from March to Dec in 2017 (10 months), to learn parameters of \emph{\model} for time series embedding in an unsupervised fashion. Then, we leverage the data from Jan 2018 to Aug 2018 to evaluate the user identification performance based on the generated embeddings from the learned model. Specifically, we perform the training/test process over the period of Jan 2018 to Aug 2018 with a sliding window of two months, \ie, (Jan $\rightarrow$ training, Feb $\rightarrow$ test);...; (Jul $\rightarrow$ training, Aug $\rightarrow$ test). The training month provides the labels to learn the classifier and the test month is used to evaluate the prediction accuracy. Irrespective of the training/ test month combination, the model parameters are learned on the basis of 2017 data in an unsupervised fashion. This allows us to also consider the generalization of performance over time. Moreover, to ensure the fairness of performance comparison, the day-long time series of users shown in the test set are not visible in the training set using our partition strategy. The training/test partition method on Fitbit data is similar as the Garmin data.

\end{itemize}

\subsection{\bf Reproducibility}
We implemented all the deep learning baselines and the proposed \emph{\model} framework with Tensorflow. For the sake of fair comparison, all experiments are conducted across all participants in the testing data and the average performance is reported. Furthermore, the validation was run ten times and the average performance numbers are reported.
In our experiments, we set the embedding dimension, the support size, positive sampling size, negative sampling size as 64, 6, 2, 4, respectively. The margin value is set as 1 and the Siamese-triplet weight is set as 4. We utilized the Glorot initialization~\cite{glorot2010understanding} and grid search~\cite{grover2016node2vec} strategies for hyperparameter initialization and tuning of all compared methods. The early stopping~\cite{raskutti2014early} is adopted to terminate the training process based on the validation performance. After the parameter tuning on all baselines, we reported their best performance in the evaluation results. During the model learning process, we used the Adam optimizer, where the batch size and learning rate were set as 64 and 0.001, respectively. \\\vspace{-0.1in}

\begin{table*}[t]
\centering
\small
\caption{User identification performance in terms of \emph{F1-score}, \emph{Accuracy} and \emph{AUC}.}
\vspace{-0.5em}
\begin{tabular}{l| c| c| c| c| c| c| c| c| c| c| c| c}
\toprule
Data & \multicolumn{12}{c}{Garmin Heart Rate Data}\\
\midrule
Month  & \multicolumn{3}{c|}{Feb} & \multicolumn{3}{c|}{Apr} & \multicolumn{3}{c|}{June} & \multicolumn{3}{c}{Aug} \\
\midrule
Method  & F1 & Acc. & AUC & F1 & Acc. & AUC & F1 & Acc. & AUC & F1 & Acc. & AUC \\
\midrule
CAE & 0.658 & 0.650 & 0.711 & 0.680 & 0.672 & 0.737 & 0.660 & 0.669 & 0.742 & 0.684& 0.698& 0.770\\
DSR & 0.649 & 0.643 & 0.692& 0.685 & 0.670 & 0.727&  0.708 & 0.708 & 0.784 & 0.694& 0.696& 0.772\\
MLR & 0.652 & 0.637 & 0.685 &0.689 & 0.670 &0.731 & 0.707 & 0.702 & 0.773 & 0.703& 0.704& 0.776\\
STN & 0.616 & 0.572 & 0.598 & 0.732 & 0.737 & 0.816 & 0.750 & 0.758 &  0.843 & 0.733& 0.740& 0.817\\
Wave2Vec & 0.743 & 0.751 & 0.832 & 0.796 & 0.803 & 0.884 & 0.816 & 0.822 &  0.901 & 0.824 & 0.830 & 0.909 \\
DeepHeart  & 0.731 & 0.738 & 0.812 & 0.753 & 0.757 & 0.834 & 0.766 & 0.773 & 0.855 & 0.750& 0.756& 0.844\\
\midrule
\model  & \textbf{0.764}  & \textbf{0.764} & \textbf{0.841}& \textbf{0.805} & \textbf{0.806} & \textbf{0.890}& \textbf{0.830} & \textbf{0.833} & \textbf{0.913}& \textbf{0.858} & \textbf{0.862} & \textbf{0.935} \\
\midrule\midrule
Data & \multicolumn{12}{c}{Fitbit Heart Rate Data}\\
\midrule
Month & \multicolumn{3}{c|}{Feb} & \multicolumn{3}{c|}{Apr} & \multicolumn{3}{c|}{June} & \multicolumn{3}{c}{Aug} \\
\midrule
Method  & F1 & Acc. & AUC & F1 & Acc. & AUC & F1 & Acc. & AUC & F1 & Acc. & AUC \\
\midrule
CAE  &0.579 & 0.577 & 0.602 &  0.596 & 0.594 & 0.641 & 0.602 & 0.587 & 0.634 & 0.606 & 0.599 & 0.638 \\
DSR & 0.647 & 0.646 &0.696 & 0.630 & 0.617  &  0.654 & 0.619 & 0.612 & 0.671 & 0.613 & 0.598 & 0.641\\
MLR & 0.643 & 0.639 & 0.697 & 0.626 & 0.624 & 0.680 & 0.6737 & 0.665 & 0.740 & 0.659 & 0.649 & 0.705 \\
STN & 0.712 & 0.718 & 0.778 & 0.687 & 0.689 & 0.749 & 0.686 & 0.681 & 0.746 & 0.729 & 0.727 & 0.787 \\
Wave2Vec & 0.778 & 0.778 & 0.862 & 0.777 & 0.777 & 0.857 & 0.771 & 0.770 & 0.857 & 0.813 & 0.812 & 0.893 \\
DeepHeart  &0.678 & 0.680 & 0.746  &0.715 & 0.718 & 0.787 & 0.719 & 0.722 & 0.791 & 0.715 & 0.714 & 0.784\\
\midrule
\model & \textbf{0.799} & \textbf{0.797} & \textbf{0.869}  & \textbf{0.804} & \textbf{0.803} & \textbf{0.886} & \textbf{0.786} & \textbf{0.782} & \textbf{0.868}& \textbf{0.833} & \textbf{0.831} & \textbf{0.914} \\
\bottomrule
\end{tabular}
\label{tab:result_user_identification}
\end{table*}
 
\begin{wraptable}{r}{7.5cm}
 \centering
 \small
 \caption{User demographic inference results. Mic-F1: Micro-F1; Mac-F1: Macro-F1}
 \vspace{-0.5em}
 \begin{tabular}{l| c| c| c| c}
 \toprule
 Demographic & \multicolumn{2}{c|}{Age} & \multicolumn{2}{c}{Gender}  \\
 \midrule
 Metric & Mic-F1 & Mac-F1 & Mic-F1 & Mac-F1 \\
 \midrule
 CAE & 0.517 & 0.364 & 0.674 & 0.655 \\
 DSR & 0.505 & 0.259 & 0.645 & 0.633 \\
 MLR & 0.523 & 0.270 & 0.645 & 0.628 \\
 STN & 0.529 & 0.346 & 0.627 & 0.622 \\
 Wave2Vec & 0.546 & 0.354 & 0.657 & 0.645 \\
 DeepHeart & 0.494 & 0.266 & 0.639 & 0.618  \\
\midrule
\model & \textbf{0.558} & \textbf{0.377} & \textbf{0.703} & \textbf{0.672} \\
 \bottomrule
 \end{tabular}
 \label{tab:result_demographic}
 \end{wraptable}

\subsection{Performance Comparison (Q1 and Q2)}

\subsubsection{\bf User Identification} Table~\ref{tab:result_user_identification} shows the user identification performance on both datasets. We can observe that \emph{\model} achieves the best performance and obtains significant improvement over state-of-the-art methods in all cases. This sheds lights on the benefit of \emph{\model},  which effectively captures the unique signatures of individuals. Although other neural network-based methods preserve the temporal structural information to learn latent representations for each individual time series, they ignore the variable-length and incompleteness of wearable-sensory data, which reveals the practical difficulties in learning accurate models across non-continuous time steps. 

To make thorough evaluation, we conduct comparison experiment of \emph{\model} and all baselines across different training and test time periods (\eg, Feb, Apr, Jun and Aug). We can note that the best performance is consistently achieved by \emph{\model} with different forecasting time periods, which reflects the robustness of \emph{\model} in learning the latent representations over time.

\subsubsection{\bf Demographic Inference}
The demographic inference performance comparison between \emph{\model} and other competitive methods on the Garmin heart rate data is shown in Table~\ref{tab:result_demographic}. We can note that our \emph{\model} outperforms other baselines in inferring users' age and gender information, which further demonstrate the efficacy of our \emph{\model} in learning significantly better time series embeddings than existing state-of-the-art methods. Similar results can be observed for the Fitbit heart rate data. In summary, the advantage of \emph{\model} lies in its proper consideration of comprehensive temporal pattern fusion for time series data.

\subsubsection{\bf Personality Prediction}
\begin{wrapfigure}{r}{7.5cm}
\vspace{-1em}
    \centering
    \subfigure[][Conscientiousness]{
        \centering
        \includegraphics[width=0.22\textwidth]{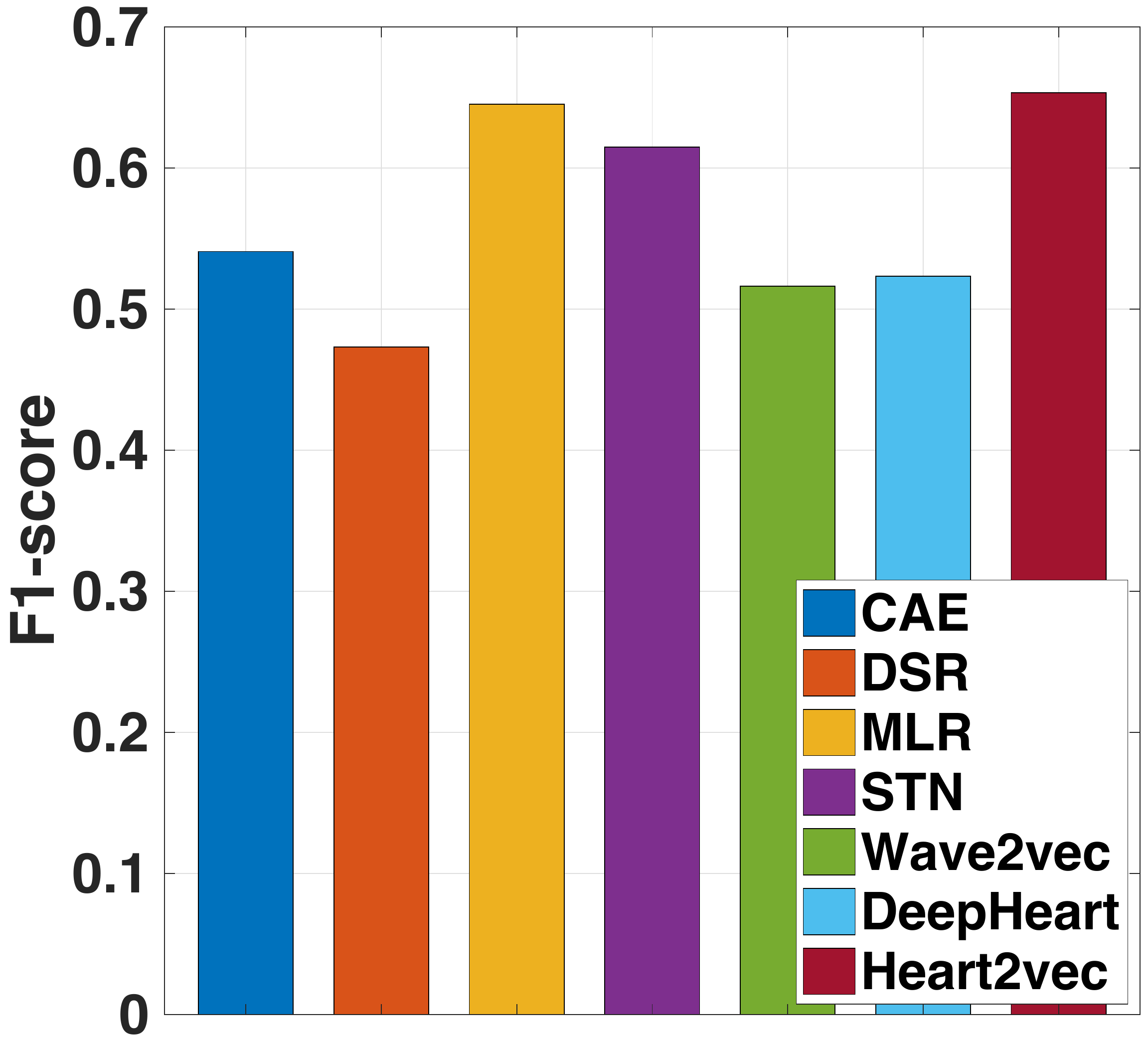}
        \label{fig:Micro}
        }
    \subfigure[][Agreeableness]{
        \centering
        \includegraphics[width=0.22\textwidth]{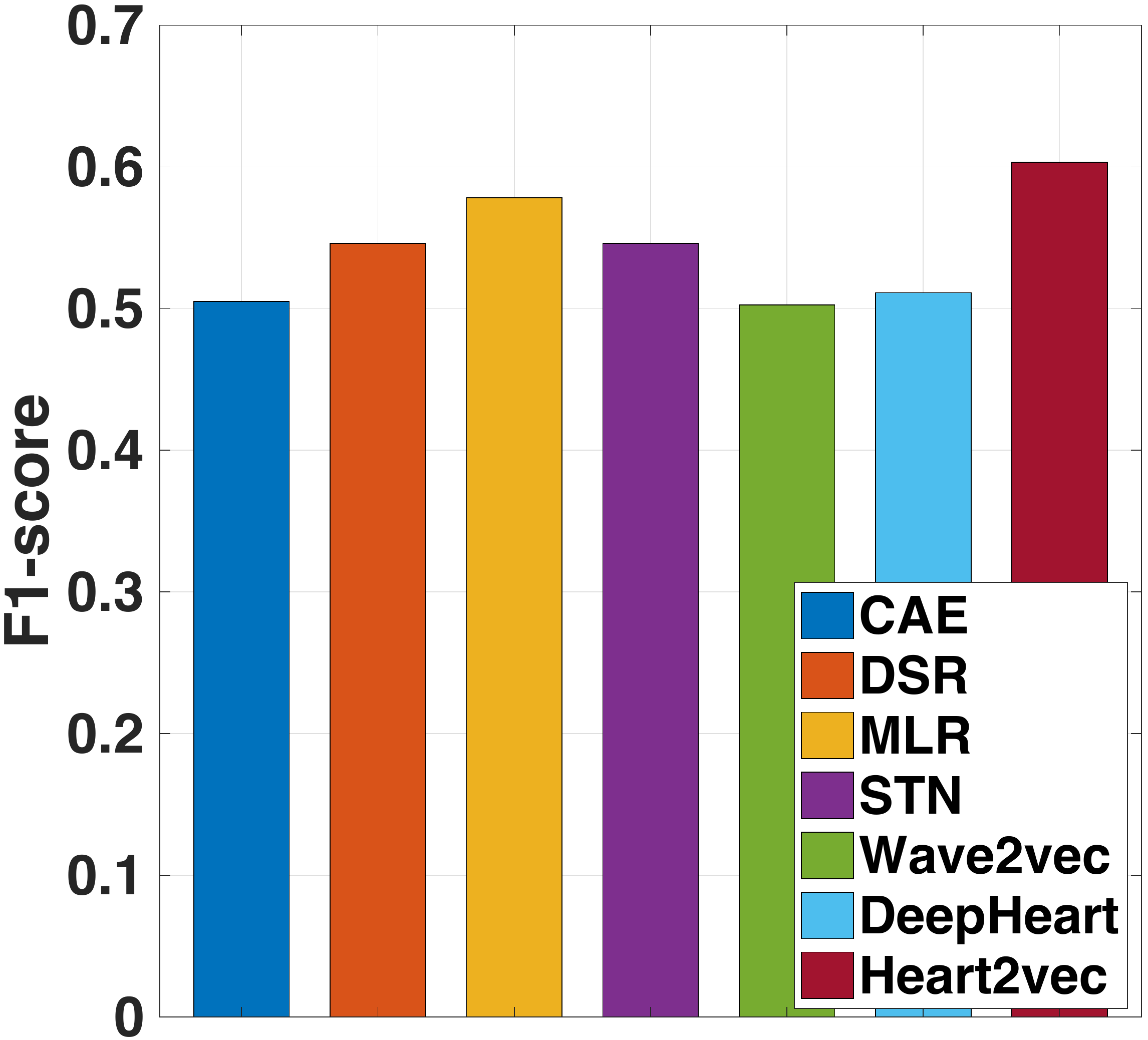}
        \label{fig:Macro}
        }
    \vspace{-0.5em}
    \caption{Personality detection results.}
    \label{fig:result_personality}
\vspace{-1.5em}
\end{wrapfigure}

Figure~\ref{fig:result_personality} shows the personality prediction results on two different categories (\ie, Agreeableness and Conscientiousness). We can observe that \emph{\model} achieves the best performance in all personality cases. The performance is followed by DSR which extracts both multi-level temporal features during the representation learning process. This further verifies the utility of temporal pattern fusion in mapping time series data into common latent space. 
\begin{wrapfigure}{r}{7.5cm}
\vspace{-1em}
    \centering
    \subfigure[][Micro-F1]{
        \centering
        \includegraphics[width=0.22\textwidth]{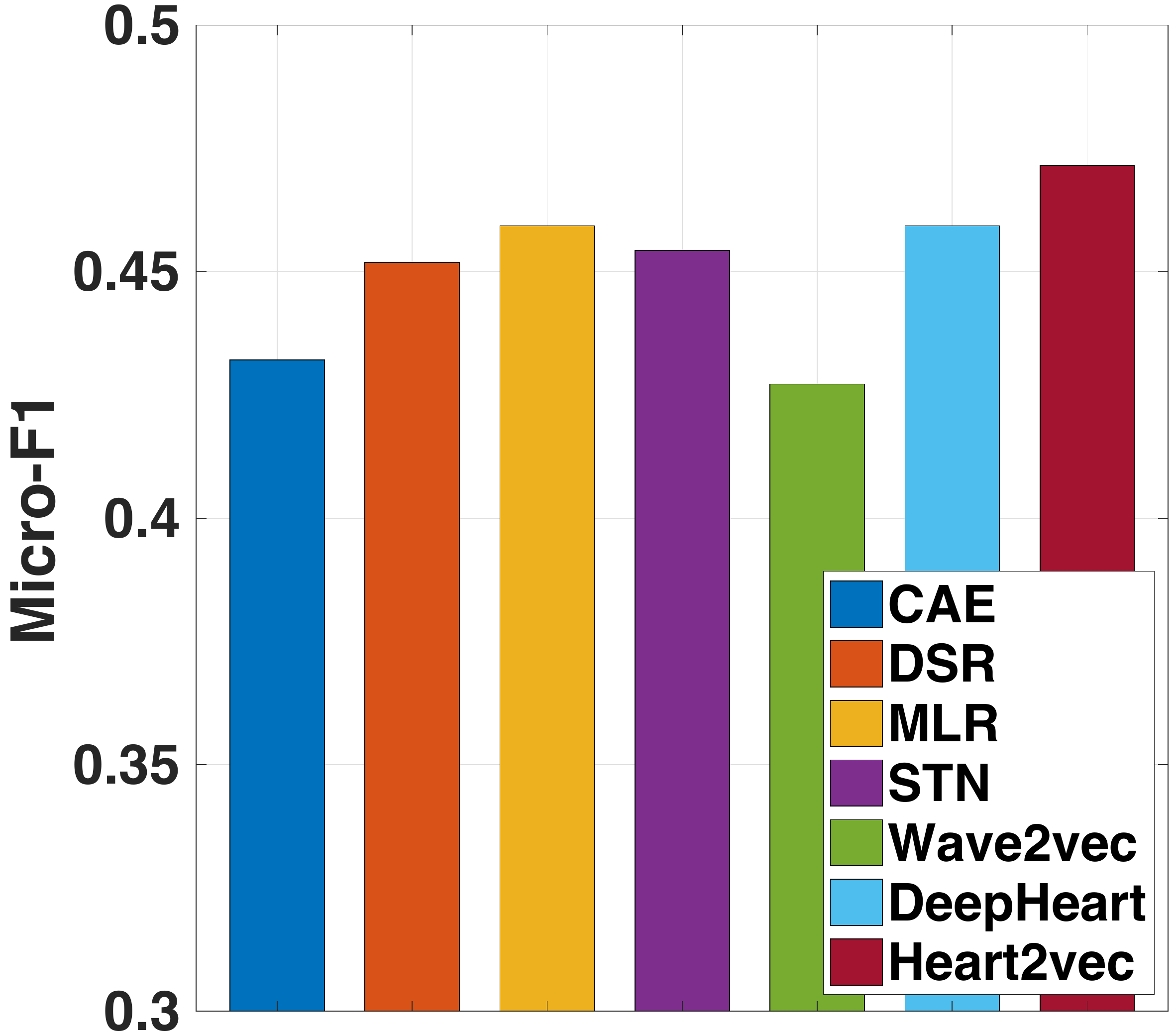}
        \label{fig:Micro}
        }
    \subfigure[][Macro-F1]{
        \centering
        \includegraphics[width=0.22\textwidth]{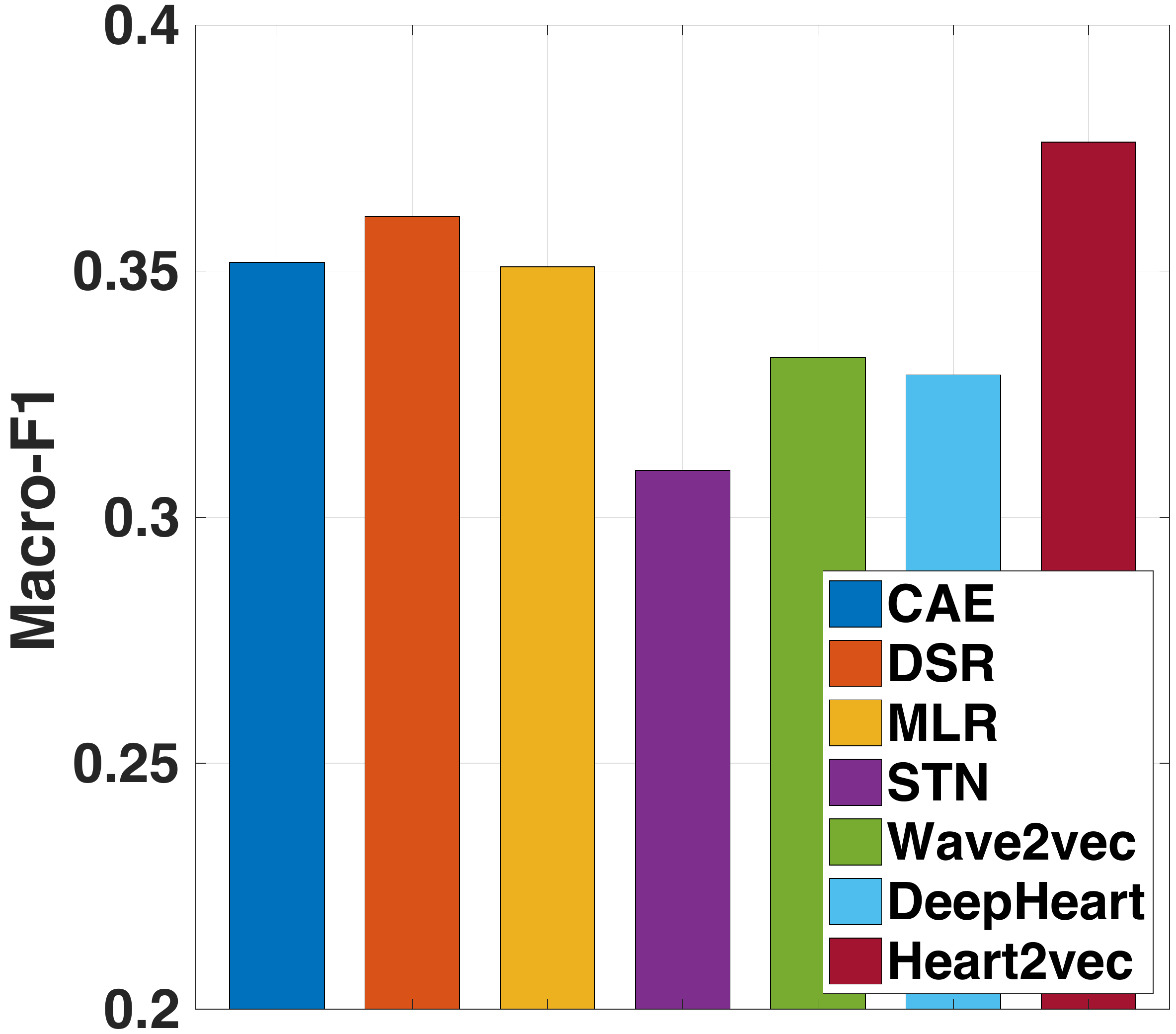}
        \label{fig:Macro}
        }
    \vspace{-0.5em}
    \caption{Job performance prediction results.}
    \label{fig:working}
    \vspace{-1.5em}
\end{wrapfigure}

\subsubsection{\bf Job Performance Prediction}
The results of job performance prediction are presented in Figure~\ref{fig:working}. In these figures, we can notice that \emph{\model} achieves the best performance in terms of Macro-F1 and Micro-F1.

\subsubsection{\bf Sleep Duration Inference} 
\begin{wrapfigure}{r}{7.5cm}
\vspace{-1em}
    \centering
    \subfigure[][MSE]{
        \centering
        \includegraphics[width=0.22\textwidth]{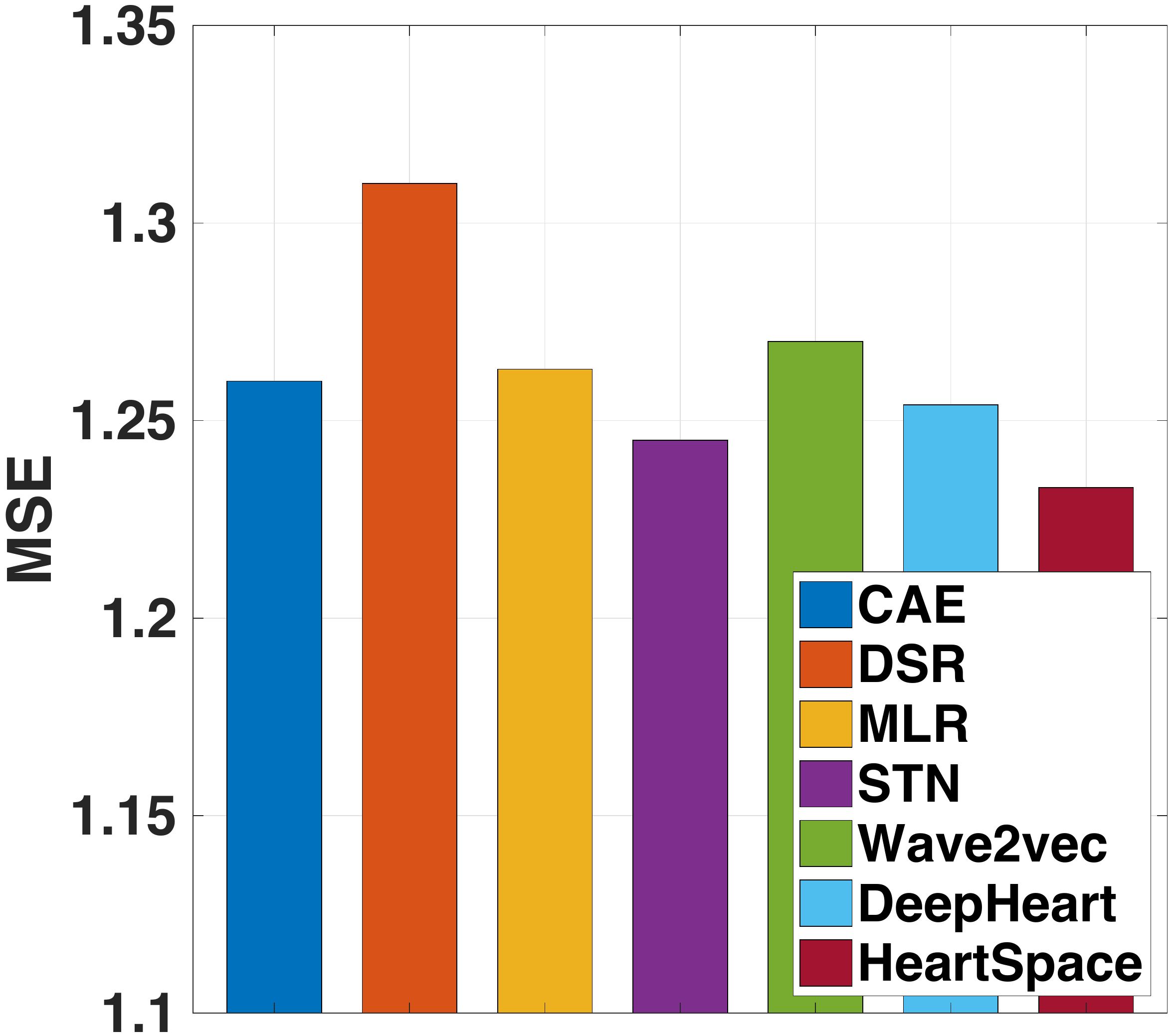}
        \label{fig:MSE}
        }
    \subfigure[][MAE]{
        \centering
        \includegraphics[width=0.22\textwidth]{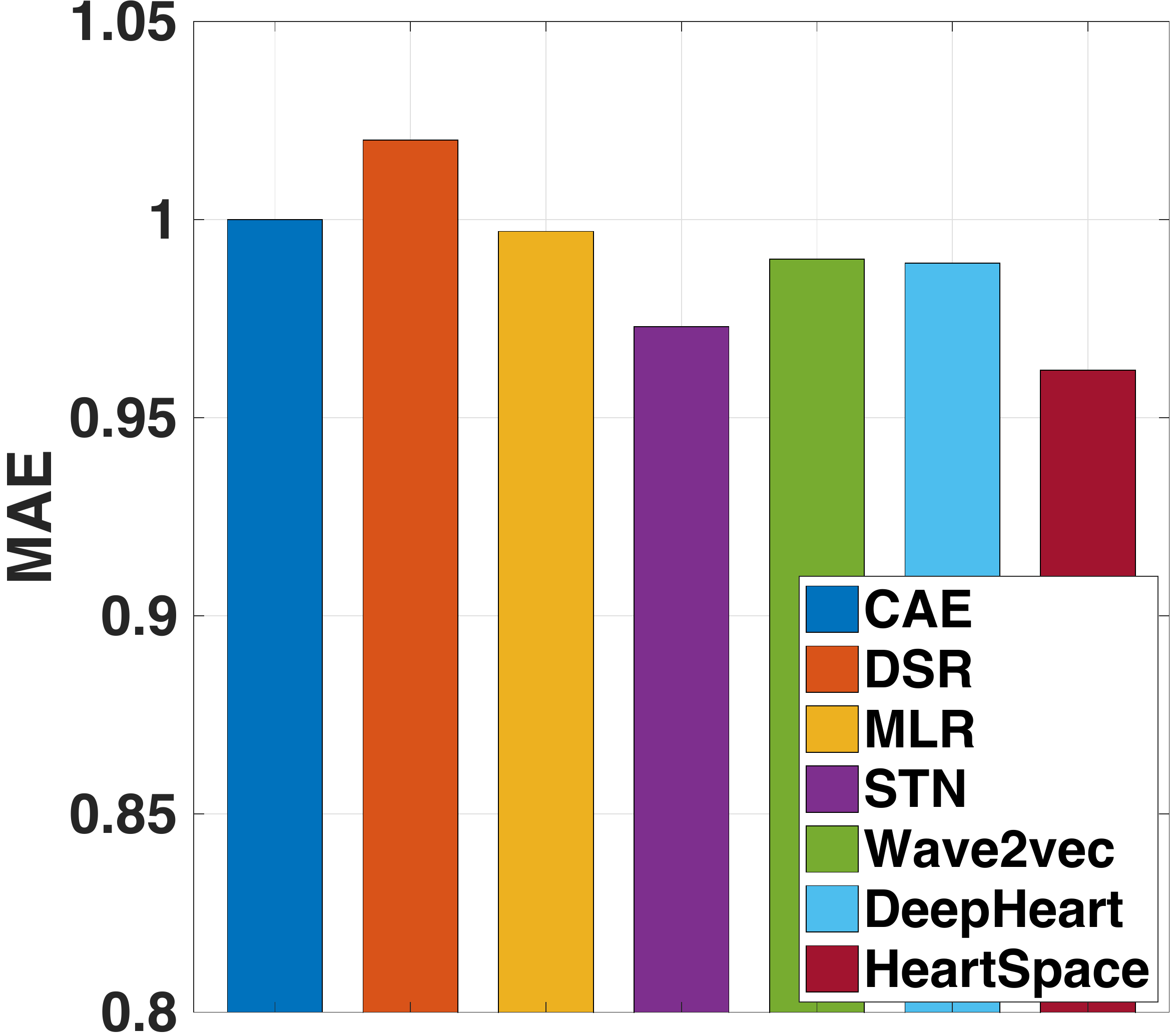}
        \label{fig:MAE}
        }
        \vspace{-1em}
    \caption{Sleep duration forecasting results.}
    \label{fig:sleep_duration}
\vspace{-1.5em}
\end{wrapfigure}

We further evaluate the proposed \emph{\model} with the application of forecasting the sleep duration of people. The performance evaluation results (measured by MSE and MAE) are presented in Figure~\ref{fig:sleep_duration}. We can observe that significant performance improvements are consistently obtained by \emph{\model} over state-of-the-art baselines, which further validates the effectiveness of \emph{\model}.

\subsection{Model Ablation: Component-Wise Evaluation of \emph{\model} (Q3)}
We also aim to get a better understanding of key components of \emph{\model}. In our evaluation, we consider three variants of the proposed method corresponding to different analytical aspects:
\begin{itemize}[leftmargin=*]
\item \textbf{Effect of Siamese-triplet Network} \emph{\model}-s. A simplified version of \emph{\model} which does not include Siamese-triplet network to model intra- and inter-time series inter-dependencies.
\item \textbf{Effect of Deep Autoencoder Module}. \emph{\model}-a. A simplified version of \emph{\model} without deep autoencoder module, \ie, only consider $\mathcal{L}_{s}$ (Siamese-triplet Network) in the loss function. 
\item \textbf{Effect of Multi-head Attention Network}. \emph{\model}-h: A variant of \emph{\model} without the multi-head attention network to learn the weights of different day-specific temporal patterns.
\end{itemize}

\begin{figure}[h]
    \centering
    \subfigure[][User Identification]{
        \centering
        \includegraphics[width=0.23\textwidth]{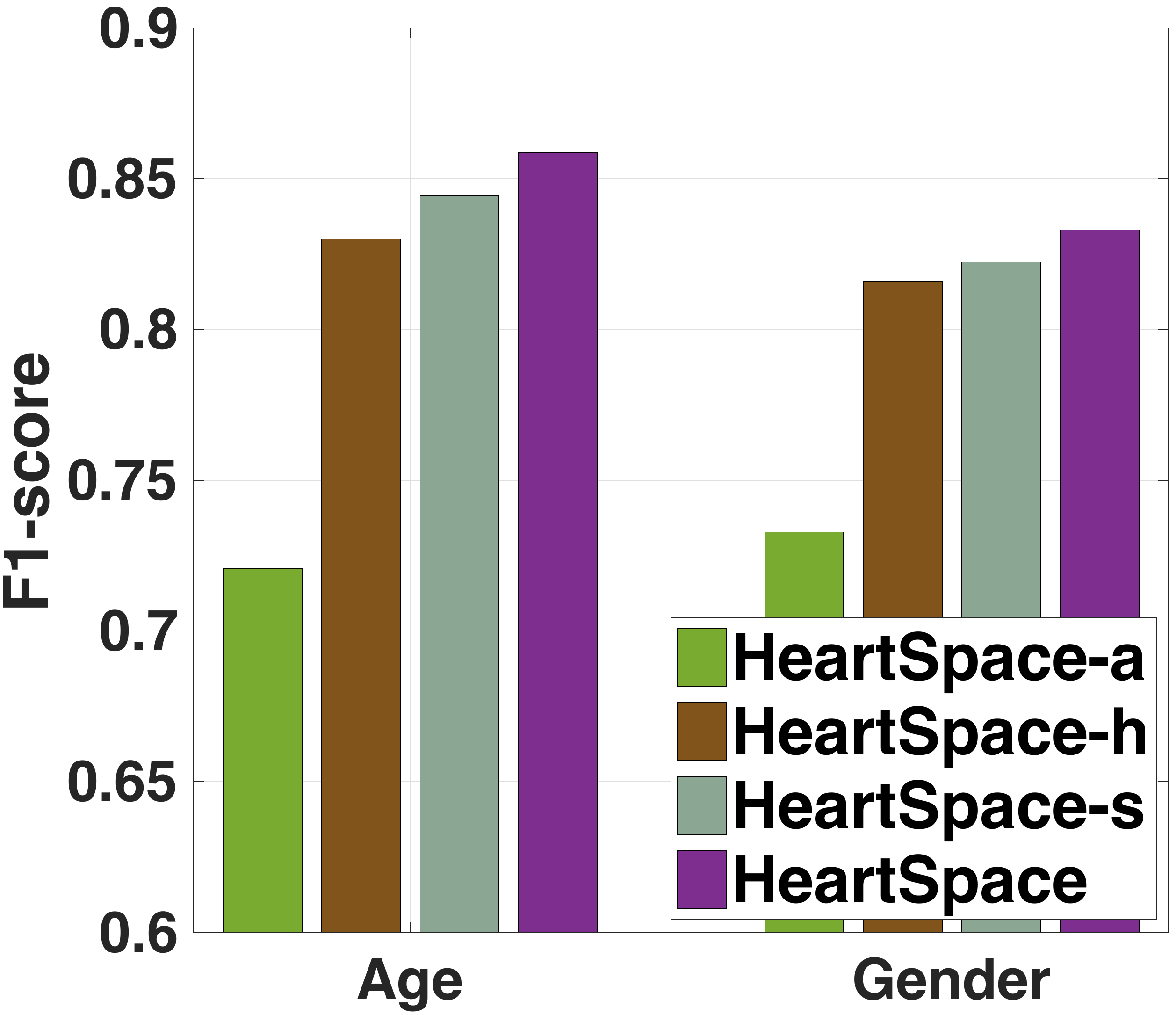}
        \label{fig:Macro}
        }
    \subfigure[][Demographic Inference]{
        \centering
        \includegraphics[width=0.23\textwidth]{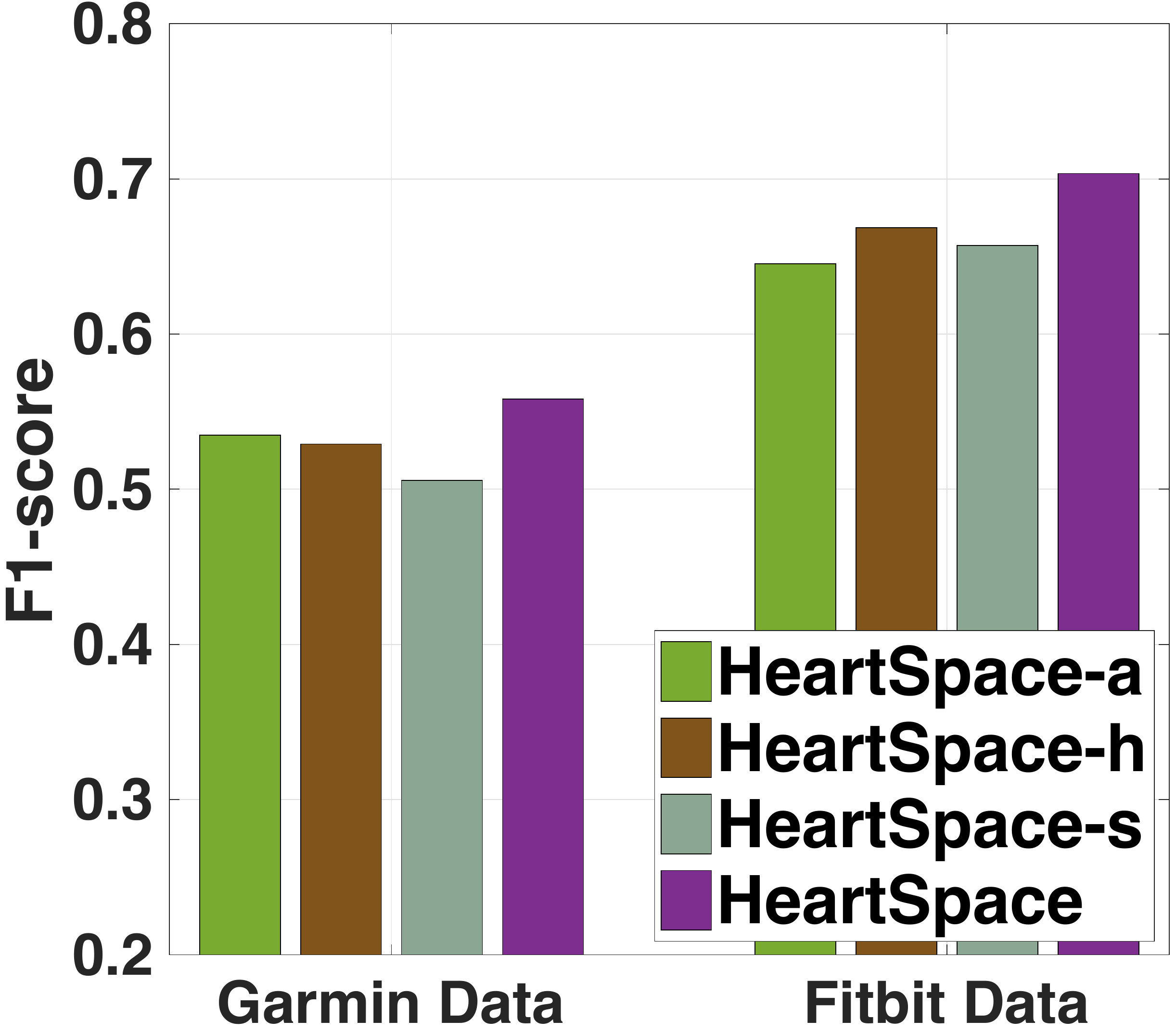}
        \label{fig:Micro}
        }
    \subfigure[][Personality Prediction]{
        \centering
        \includegraphics[width=0.23\textwidth]{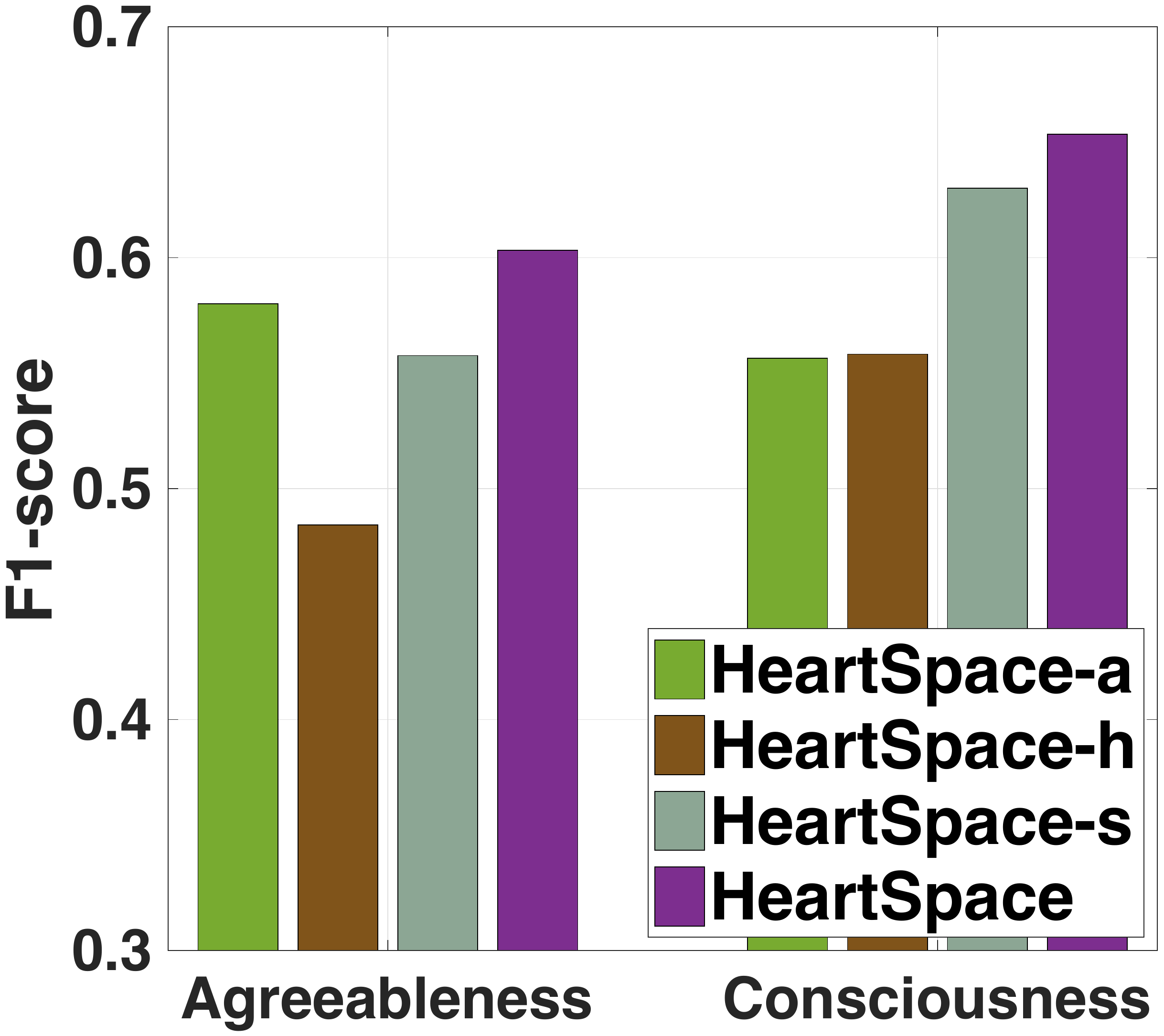}
        \label{fig:Micro}
        }
    \subfigure[][Job Performance Prediction]{
        \centering
        \includegraphics[width=0.23\textwidth]{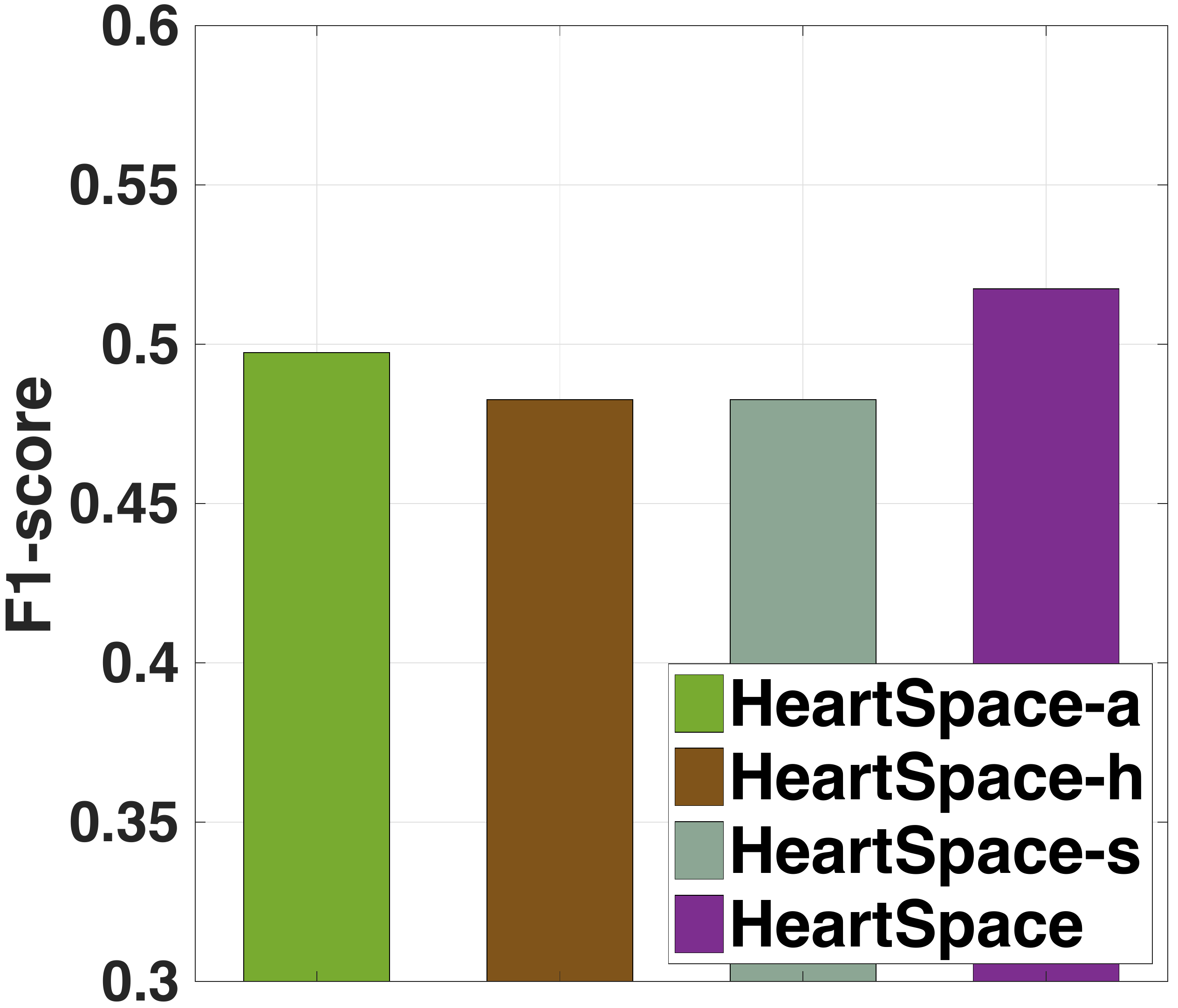}
        \label{fig:Micro}
        }
    \vspace{-0.5em}
    \caption{Evaluation on \emph{\model} variants.}
    \label{fig:model_variants}
\end{figure}

We report the results in Figure~\ref{fig:model_variants}. Notice that the full version of \emph{\model} achieves the best performance in all cases, which suggests: (i) The efficacy of the designed Siamese-triplet network optimization strategy for preserving structural information of implicit intra- and inter-time series correlations. (ii) The effectiveness of \emph{\model} in capturing complex temporal dependencies across time steps for variable-length sensor data. (iii) The effectiveness of \emph{\model} in exploring feature modeling in different representation spaces during our pattern fusion process. As such, it is necessary to build a joint framework to capture multi-dimensional correlations in wearable-sensory time series representation learning.

\begin{figure}[h]
    \centering
    \subfigure[][DeepHeart]{
        \centering
        \includegraphics[width=0.35\textwidth]{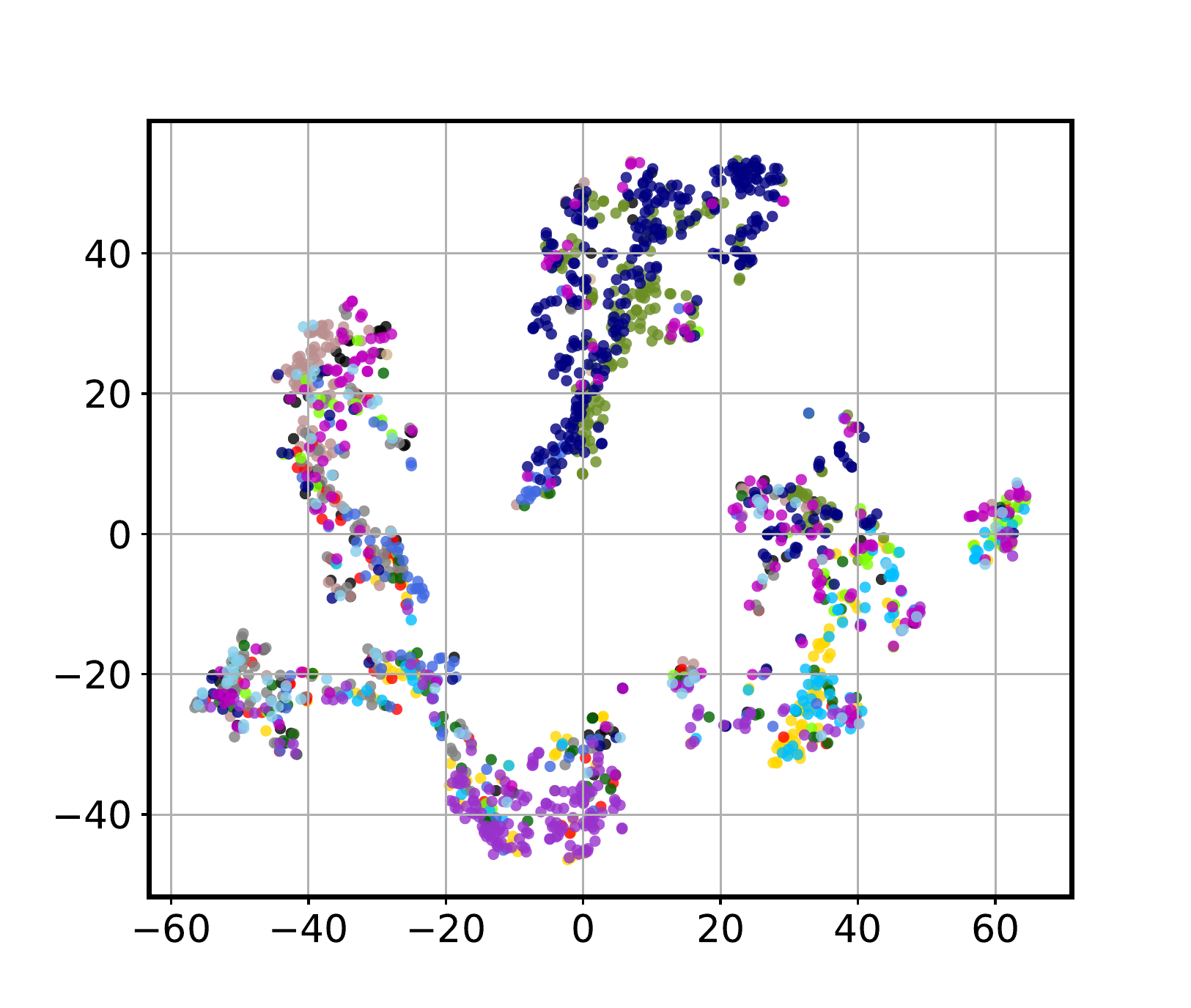}
        \label{fig:Macro}
        }
    \subfigure[][\model]{
        \centering
        \includegraphics[width=0.35\textwidth]{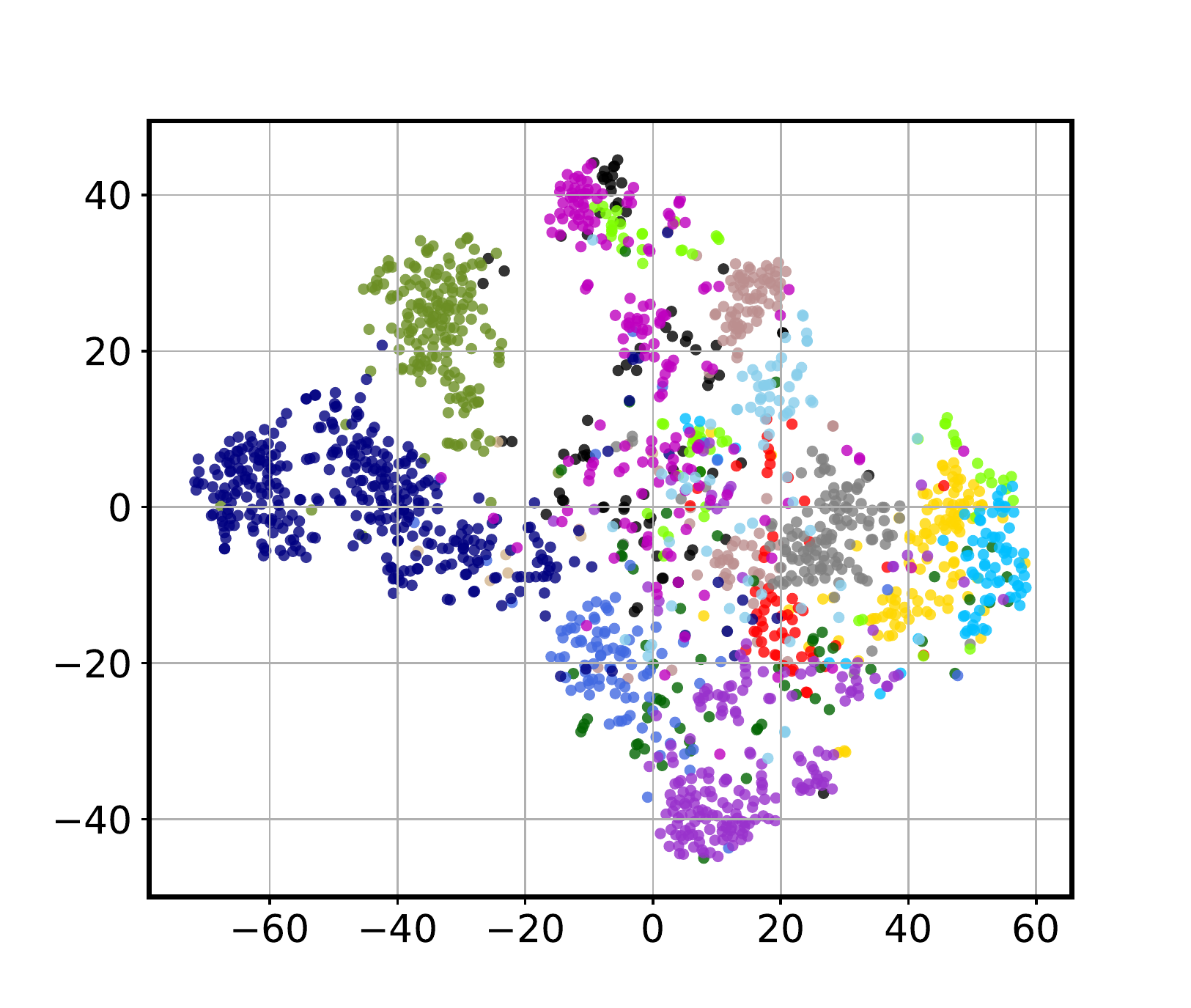}
        \label{fig:Micro}
        }
    \caption{Embedding visualizations of \model\ v.s. DeepHeart. 2D t-SNE projections of the 64D embeddings of all day-specific time series from 15 participants on the Garmin heart rate data. Each color indicates an individual user.}
    \label{fig:case_study}
\end{figure}

\subsection{\bf Hyperparameter Studies (Q4)}
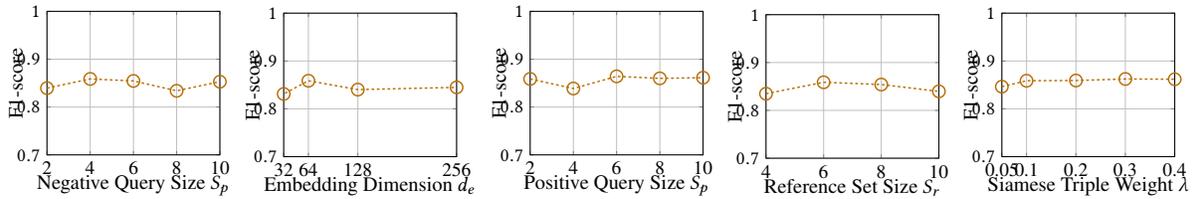
\begin{figure}[h]
    \centering
        \begin{adjustbox}{max width=\linewidth}
            \input{fig/parameter_sensitivity_overall}
        \end{adjustbox}
        \caption{Hyperparameter Studies in terms of \emph{F1-score}.}
        \label{fig:parameter_study_overall}
\end{figure}

To investigate the robustness of \emph{\model}, we examined how the different choices of five key parameters affect the performance of \emph{\model}. Figure~\ref{fig:parameter_study_overall} shows the evaluation results as a function of one selected parameter when fixing others. Overall, we observe that \emph{\model} is not strictly sensitive to these parameters and is able to reach high performance under a cost-effective parameter choice, which demonstrates the robustness of \emph{\model}.

Furthermore, we can observe that the increase of prediction performance saturates as the representation dimensionality increases. This is because: at the beginning, a larger value of embedding dimension brings a stronger representation power for the recent framework, but the further increase of dimension size of latent representations might lead to the overfitting issue. In our experiments, we set the dimension size as 64 due to the consideration of the performance and computational cost. We can observe that both the positive query set size and negative query set size, as well as siamese triple loss coefficient have a relatively low impact on the model performance.

\subsection{Case Study: Visualization (Q5)}
We employ the TensorFlow embedding projector to further visualize the low-dimensional time series representations learned by  \emph{\model} and one selected baseline (DeepHeart) on Garmin heart rate dataset. Figure~\ref{fig:case_study} shows the visualization of day-specific time series embeddings from randomly selected 15 users. We can observe that the embeddings from the same user could be identified by our \emph{\model} and cluster them closer than other embeddings, while the embeddings learned by DeepHeart of different users are mixed and cannot be well identified. Therefore, our model generates more accurate feature representations of user's time series data and is capable of preserving the unique signatures of each individual, which can then be leveraged for similarity analyses. 

%% file: fig/parameter_sensitivity_overall.tex
\begin{tikzpicture}
\begin{axis}[
    xlabel={Negative Query Size $S_p$},
    ylabel={F1-score},
    xmin=2,xmax=10,
    ymin=0.7,ymax=1.0,
    xtick={2,4,6,8,10},
    xticklabels={2,4,6,8,10},
    legend cell align=right,
    grid=both,
    ylabel style={align=center},
    every axis plot/.append style={ultra thick},
    every tick label/.append style={scale=2.15},
    label style={scale=2.3},
    legend style={
        nodes={scale=1.5, transform shape},
        legend image post style={scale=1.5},
        },
    legend style={at={(0,0)},anchor=south west},
    every axis plot post/.append style={
        every mark/.append style={scale=3.5}
    }
]

\addplot[color={rgb:red,4;green,2;yellow,1}, mark=o, dashed, mark options={solid}]
    table[x=para, y=f1] {num_negative_f1.txt};

\end{axis}
\end{tikzpicture}

\begin{tikzpicture}
\begin{axis}[
    xlabel={Embedding Dimension $d_e$},
    ylabel={F1-score},
    xmin=32,xmax=256,
    ymin=0.7,ymax=1.0,
    xtick={32,64,128,256},
    xticklabels={32,64,128,256},
    legend cell align=right,
    grid=both,
    ylabel style={align=center},
    every axis plot/.append style={ultra thick},
    every tick label/.append style={scale=2.15},
    label style={scale=2.3},
    legend style={
        nodes={scale=1.5, transform shape},
        legend image post style={scale=1.5},
        },
    legend style={at={(0,0)},anchor=south west},
    every axis plot post/.append style={
        every mark/.append style={scale=3.5}
    }
]

\addplot[color={rgb:red,4;green,2;yellow,1}, mark=o, dashed, mark options={solid}]
    table[x=para, y=f1] {embedding_dimension_f1.txt};

\end{axis}
\end{tikzpicture}

\begin{tikzpicture}
\begin{axis}[
    xlabel={Positive Query Size $S_p$},
    ylabel={F1-score},
    xmin=2,xmax=10,
    ymin=0.7,ymax=1.0,
    xtick={2,4,6,8,10},
    xticklabels={2,4,6,8,10},
    legend cell align=right,
    grid=both,
    ylabel style={align=center},
    every axis plot/.append style={ultra thick},
    every tick label/.append style={scale=2.15},
    label style={scale=2.3},
    legend style={
        nodes={scale=1.5, transform shape},
        legend image post style={scale=1.5},
        },
    legend style={at={(0,0)},anchor=south west},
    every axis plot post/.append style={
        every mark/.append style={scale=3.5}
    }
]

\addplot[color={rgb:red,4;green,2;yellow,1}, mark=o, dashed, mark options={solid}]
    table[x=para, y=f1] {num_positive_f1.txt};

\end{axis}
\end{tikzpicture}

\begin{tikzpicture}
\begin{axis}[
    xlabel={Reference Set Size $S_r$},
    ylabel={F1-score},
    xmin=4,xmax=10,
    ymin=0.7,ymax=1.0,
    xtick={4,6,8,10},
    xticklabels={4,6,8,10},
    legend cell align=right,
    grid=both,
    ylabel style={align=center},
    every axis plot/.append style={ultra thick},
    every tick label/.append style={scale=2.15},
    label style={scale=2.3},
    legend style={
        nodes={scale=1.5, transform shape},
        legend image post style={scale=1.5},
        },
    legend style={at={(0,0)},anchor=south west},
    every axis plot post/.append style={
        every mark/.append style={scale=3.5}
    }
]

\addplot[color={rgb:red,4;green,2;yellow,1}, mark=o, dashed, mark options={solid}]
    table[x=para, y=f1] {num_reference_f1.txt};

\end{axis}
\end{tikzpicture}

\begin{tikzpicture}
\begin{axis}[
    xlabel={Siamese Triple Weight $\lambda$},
    ylabel={F1-score},
    xmin=0.05,xmax=0.4,
    ymin=0.7,ymax=1.0,
    xtick={0.05,0.1,0.2,0.3,0.4},
    xticklabels={0.05,0.1,0.2,0.3,0.4},
    legend cell align=right,
    grid=both,
    ylabel style={align=center},
    every axis plot/.append style={ultra thick},
    every tick label/.append style={scale=2.15},
    label style={scale=2.3},
    legend style={
        nodes={scale=1.5, transform shape},
        legend image post style={scale=1.5},
        },
    legend style={at={(0,0)},anchor=south west},
    every axis plot post/.append style={
        every mark/.append style={scale=3.5}
    }
]

\addplot[color={rgb:red,4;green,2;yellow,1}, mark=o, dashed, mark options={solid}]
    table[x=para, y=f1] {triplet_loss_weight.txt};

\end{axis}
\end{tikzpicture}

%% file: relate.tex
\section{Related Work}
\label{sec:relate}




\noindent \textbf{Representation Learning Models}.
With the advent of deep learning techniques, significant effort has been devoted to developing neural network-based representation learning models on various data. For the word representation learning, there exist a good amount of literature on designing techniques by considering word context in a document, such as word2vec~\cite{mikolov2013distributed} and Glove~\cite{pennington2014glove}. Many follow-up works extend the basic framework to learn latent representations on multimedia data~\cite{ramanathan2015learning,goroshin2015unsupervised} and network data~\cite{tu2018deep,grover2016node2vec}, in order to capture frame-level temporal dependency and network structural information, respectively. For example, node2vec~\cite{grover2016node2vec} proposes to learn node embeddings based on random walk. However, the wearable-sensory time series representation learning remains to be a critical but largely unsolved question. This work proposes a principled framework to address this challenge by automatically preserving the underlying
structure of dynamic temporal patterns of wearable-sensory time series data. \\\vspace{-0.1in}



\noindent \textbf{Deep Learning for Healthcare Applications}. With the advent of deep learning techniques, many deep neural network frameworks have been proposed to address various challenges in healthcare informatics~\cite{bai2018interpretable,ma2018risk,cao2017deepmood,suhara2017deepmood,choi2017gram}. For example, Cao~\etal \cite{cao2017deepmood} proposed an end-to-end deep architecture for the prediction of human mood. Bai~\etal \cite{bai2018interpretable} modeled sequential patient data and predict future clinical events. This work furthers this direction of investigation by proposing a general time series representation learning framework to capture hierarchical structural correlations exhibited from human heart rate data, which cannot be handled by previous models. \\\vspace{-0.1in}


\noindent \textbf{Representation Learning on wearable-sensory Data}. A handful of studies~\cite{ballinger2018deepheart,amiriparian2018deep,lehman2018representation} have investigated the representation learning on wearable-sensory data. Ballinger~\etal \cite{ballinger2018deepheart} proposed a semi-supervised learning method to detect health conditions. Amiriparian~\etal \cite{amiriparian2018deep} explored the deep feature representations to aid the diagnosis of cardiovascular disorders with sequence to sequence autoencoders. While these pioneering studies have demonstrated the effectiveness of representation learning on wearable-sensory data, they do not address our problem of designing a unified model that dynamically captures the evolving temporal characteristics from wearable-sensory time series with variable-length.

%% file: conclusion.tex
\section{Conclusion}
\label{sec:conclusion}

In this paper, we presented  \emph{\model}, a novel time series representation learning method for wearable-sensory data that addresses several challenges stemming from such data and also overcomes the limitations of current state-of-the-art approaches, including dealing with incomplete and variable length time series, intra-sensor / individual variability, and absence of ground truth.  \emph{\model} first learns latent representation to encode temporal patterns of individual day-specific time series with a deep autoencoder model. Then, an integrative framework of a pattern aggregation network and a Siamese-triplet network optimization maps  variable-length wearable-sensory time series into the common latent space such that the implicit intra-series and inter-series correlations well preserved. Extensive experiments on real-world data, representing different sensor and subjects' distribution,  demonstrate that the latent feature representations learned by \emph{\model} are significantly more accurate and generalizable than the contemporary methods. 


 
 